\documentclass[fleqn,usenatbib]{mnras}

\usepackage{newtxtext,newtxmath}
\usepackage[OT2,T1]{fontenc}

\DeclareRobustCommand{\VAN}[3]{#2}
\let\VANthebibliography\thebibliography
\def\thebibliography{\DeclareRobustCommand{\VAN}[3]{##3}\VANthebibliography}

\usepackage{graphicx}	
\usepackage{amsmath}	
\usepackage{color}
\usepackage[dvipsnames]{xcolor}
\usepackage{booktabs}
\usepackage{tcolorbox}

\DeclareSymbolFont{cyrletters}{OT2}{wncyr}{m}{n}
\DeclareMathSymbol{\Sha}{\mathalpha}{cyrletters}{"58}

\newcommand{\umag}{\upsilon}

\newcommand{\paperitablelayers}{1}
\newcommand{\paperitableoutpsf}{4}
\newcommand{\paperifigcoveragemap}{1}
\newcommand{\paperifignoiseonef}{3}
\newcommand{\paperifiginterestingobjects}{8}
\newcommand{\paperifigtilefig}{4}
\newcommand{\paperifigfidelitymap}{10}
\newcommand{\paperifigurefcum}{11}
\newcommand{\paperifigyresfig}{7}
\newcommand{\paperisecproblem}{2}
\newcommand{\paperisecinput}{3}
\newcommand{\paperisecnoise}{3.4}
\newcommand{\paperiseccoadd}{4}
\newcommand{\paperiappsampling}{C}

\title[Image combination for Roman II]{Simulating image coaddition with the {\textit{\textbf{Nancy Grace Roman Space Telescope}}}: II. Analysis of the simulated images and implications for weak lensing}

\author[Yamamoto et al.]{Masaya Yamamoto$^1$,
Katherine Laliotis$^{2,3}$,
Emily Macbeth$^{2,3,4}$,
Tianqing Zhang$^5$,
\newauthor{Christopher M. Hirata$^{2,3,4}$, M.A.~Troxel$^1$,
Kaili Cao$^{2,3}$,
Ami Choi$^6$,
Jahmour Givans$^{7,8}$,
Katrin Heitmann$^9$,
}
\newauthor{Mustapha Ishak$^{10}$,
Mike Jarvis$^{11}$,
Eve Kovacs$^{9}$,
Heyang Long$^{2,3}$,
Rachel Mandelbaum$^{5,12}$,
Andy Park$^5$,
}
\newauthor{Anna Porredon$^{13}$,
Christopher W. Walter$^1$,
W. Michael Wood-Vasey$^{14}$}\\
$^{1}$Department of Physics, Duke University, Box 90305, Durham, NC 27708, USA\\
$^{2}$Center for Cosmology and AstroParticle Physics (CCAPP), 191 West Woodruff Ave, The Ohio State University, Columbus, OH 43210, USA\\
$^{3}$Department of Physics, The Ohio State University, 191 West Woodruff Ave, Columbus, OH 43210, USA\\
$^{4}$Department of Astronomy, The Ohio State University, 140 West 18th Avenue, Columbus, OH 43210, USA\\
$^{5}$McWilliams Center for Cosmology, Department of Physics, Carnegie Mellon University, 5000 Forbes Ave, Pittsburgh, PA 15213, USA\\
$^{6}$NASA Goddard Space Flight Center, Mail Code 665, Greenbelt, MD 20771, USA\\
$^{7}$Department of Astrophysical Sciences, Princeton University, 4 Ivy Lane, Princeton, NJ 08540, USA\\
$^{8}$Center for Computational Astrophysics, Flatiron Institute, 162 5th Ave, New York, NY 10010, USA\\
$^{9}$Argonne National Laboratory, 9700 S Cass Ave, Lemont, IL 60439, USA\\
$^{10}$Department of Physics, University of Texas at Dallas, EC36, 800 W Campbell Rd, Richardson, TX 75080, USA\\
$^{11}$Department of Physics and Astronomy, University of Pennsylvania, 209 South 33rd Street, Philadelphia, PA 19104-6396, USA\\
$^{12}$NSF AI Planning Institute for Data-Driven Discovery in Physics, Carnegie Mellon University, Pittsburgh, PA 15213, USA\\
$^{13}$Institute for Astronomy, University of Edinburgh, Edinburgh EH9 3HJ, UK\\
$^{14}$Physics and Astronomy Department, University of Pittsburgh, 4200 Fifth Ave, Pittsburgh, PA 15260, USA
}

\date{Accepted XXX. Received YYY; in original form ZZZ}

\pubyear{2023}

\begin{document}
\label{firstpage}
\pagerange{\pageref{firstpage}--\pageref{lastpage}}
\maketitle

\begin{abstract}
One challenge for applying current weak lensing analysis tools to the {\slshape Nancy Grace Roman Space Telescope} is that individual images will be undersampled. Our companion paper presented an initial application of {\sc Imcom} --- an algorithm that builds an optimal mapping from input to output pixels to reconstruct a fully sampled combined image --- on the {\slshape Roman} image simulations. In this paper, we measure the output noise power spectra, identify the sources of the major features in the power spectra, and show that simple analytic models that ignore sampling effects underestimate the power spectra of the coadded noise images. We compute the moments of both idealized injected stars and fully simulated stars in the coadded images, and their 1- and 2-point statistics. We show that the idealized injected stars have root-mean-square ellipticity errors $(1-6)\times 10^{-4}$ per component depending on the band; the correlation functions are $\ge 2$ orders of magnitude below requirements, indicating that the image combination step itself is using a small fraction of the overall {\slshape Roman} 2nd moment error budget, although the 4th moments are larger and warrant further investigation. The stars in the simulated sky images, which include blending and chromaticity effects, have correlation functions near the requirement level (and below the requirement level in a wide-band image constructed by stacking all 4 filters). We evaluate the noise-induced biases in the ellipticities of injected stars, and explain the resulting trends with an analytical model. We conclude by enumerating the next steps in developing an image coaddition pipeline for {\slshape Roman}.
\end{abstract}


\section{Introduction}

Weak gravitational lensing has long been recognized as a powerful probe to infer matter distribution and matter content in the Universe \citep[e.g.,][]{2002NewAR..46..767H, 2006astro.ph..9591A, 2013PhR...530...87W}. It has received great attention as a test of dark energy and modified gravity models, and is a leading tool in testing the consistency of $\Lambda$-CDM model between early-time (e.g., cosmic microwave background; \citealt{2020A&A...641A...6P}) and late-time (e.g., weak lensing) observations \citep{2021MNRAS.505.6179L}. Further improvement when the next-generation observatories Vera C. Rubin Observatory \citep{2009arXiv0912.0201L, 2019ApJ...873..111I}, {\slshape Euclid} \citep{2011arXiv1110.3193L} and the {\slshape Nancy Grace Roman Space Telescope} \citep{2015arXiv150303757S} start operating relies on our capability to accurately process raw images and analyze the processed images. Recent efforts to minimize the weak lensing shear systematics were mainly proposed and applied in large optical imaging survey collaborations such as the Dark Energy Survey\footnote{\url{ https://www.darkenergysurvey.org}}, Hyper-Suprime Cam\footnote{\url{https://hsc.mtk.nao.ac.jp/ssp/}} and Kilo-Degree Survey\footnote{\url{ https://kids.strw.leidenuniv.nl}}. These include a new image coaddition technique (e.g., Becker et al. in prep.) to avoid PSF discontinuity, accurate PSF modeling (e.g., \citealt{2021MNRAS.501.1282J}) and shear calibration and characterization \citep[e.g.,][]{2020ApJ...902..138S, 2022arXiv220810522L}. These are mainly developed to mitigate various biases related to cosmic shear measurement using ground-based telescopes. Images from space-based telescopes, however, are often under-sampled at the native plate scale, which introduces additional complications into precision measurements such as galaxy shapes.

Since it is not possible to recover the original astronomical scene from under-sampled images through operations such as deconvolution, object shapes measured in under-sampled images can suffer biases (see, e.g., \citealt{2023arXiv230107725F} for a study with the {\slshape Hubble Space Telescope}).
A study of the fundamental mathematical issues can be found in Appendix~\paperiappsampling\ of our companion paper (Hirata et al. in prep.; hereafter ``Paper I''). \citet{2021MNRAS.502.4048K} and \citet{2023MNRAS.519.4241Y} measured galaxy shapes with the {\sc Metacalibration} technique \citep{2017arXiv170202600H, 2017ApJ...841...24S} on simulated \emph{Euclid} or \emph{Roman} images and presented a few percent shear calibration bias. Even though these studies did not particularly identify the sources of other biases except aliasing bias, in order to meet the systematic error budget on shear calibration defined by the Science Requirements Document (SRD)\footnote{The weak lensing requirements derived during the Concept and Technology Development phase (Phase A) are described in \citet{2018arXiv180403628D}. An updated version of the SRD can be found at \url{https://asd.gsfc.nasa.gov/romancaa/docs2/RST-SYS-REQ-0020C_SRD.docx}.} we need to mitigate this aliasing bias at the image level by reconstructing well-sampled images with survey-specific dithering strategy (e.g., telescope rotation and pointing) and realistic simulated PSFs.

As discussed in \citet{2023OJAp....6E...5M}, in order for a reconstructed image to have a well-defined PSF the image coaddition algorithm needs to be linear and the weights on each exposure need to be independent of signals of sky scene (e.g., inverse-variance weights including source Poisson noise break the linearity assumption). To produce a well-defined coadded PSF, \citet{2011ApJ...741...46R} presented the linear image combination algorithm {\sc Imcom}. {\sc Imcom} finds an optimal matrix ${\mathbfss T}$ mapping from input (native) to output (coadd) pixels by minimizing the cost function which consists of the ``leakage'' (squared $L^2$ norm of the difference between the output PSF compared to a user-specified ``target'' PSF) and the output noise variance. The relative weight of the leakage and noise in the objective is controlled by a Lagrange multiplier $\kappa$ in {\sc Imcom}; the logic in the iterative solver for $\kappa$ can be configured for different outcomes (e.g., to minimize noise subject to leakage being less than a specified value).

{\sc Imcom} is fundamentally different from most other image reconstruction techniques in the sense that users are able to design their desired PSF for a given area and noise correlation level. Qualitatively, it is a process that transforms the input images with PSFs into a combined image with a common PSF in every pixel for a given area. If one would like a reconstructed image to be well-sampled and to have an isotropic and homogeneous PSF (indeed the case that is explored in Paper~I to avoid PSFs with diffraction spikes), the size of the output PSF needs to be larger than the original PSF; hence the reconstructed image would be lower resolution than a "typical" coadd image. Despite this trade-off, the size of the output PSF is still much smaller than that of a ground-based PSF (see Table~\paperitableoutpsf\ of Paper~I), meaning that this method still takes advantage of space-based imaging. Readers may refer to \citet{2011ApJ...741...46R} and Sec.~\paperisecproblem\ of Paper~I for the full details of the algorithm and the motivation in the current context of {\slshape Roman}, respectively.

With the recent development of realistic image simulations \citep{2021MNRAS.501.2044T, 2022arXiv220906829T} for \emph{Roman} following the specifications of the telescope, Paper~I re-implemented {\sc Imcom} with a \textsc{python} interface and \textsc{C} back-end so that the pipeline is compatible with the image simulations. Paper~I describes the implementation of {\sc Imcom} and updates from the original version (e.g., how to compute input PSF correlations. see Sec.~\paperiseccoadd\ of Paper~I). It then applied {\sc Imcom} on various input layers of 48 $\times$ 48 arcmin footprint (including realistic \emph{Roman} single-exposure images produced in \cite{2022arXiv220906829T}; the description of different input layers is in Sec.~\paperisecinput\ and Table~\paperitablelayers\ of Paper~I). The target PSF specified in Paper~I is the Airy disk convolved with a Gaussian kernel whose width depends on the bandpass (Paper~I Table~\paperitableoutpsf), and {\sc Imcom} is configured to produce (if possible) an output PSF within 0.1\% of the target PSF in an $L^2$ norm sense. This implies that the root-sum-square of the error in the moments is 0.1\% if the moments are defined in an orthonormal basis such as shapelets \citep{2003MNRAS.338...35R}. Paper~I demonstrates several aspects of {\sc Imcom} on {\slshape Roman} simulations:
\begin{itemize}
    \item reconstruction of Nyquist-sampled images;
    \item homogenization and isotropization of well-defined co-add PSF; and
    \item minimization of noise covariances in coadded images.
\end{itemize}
This paper goes further and examines the moments of simulated sources and the correlation functions and power spectra of the output noise and residual systematics in the coadded images. These studies are needed to support error budgeting for the {\slshape Roman} weak lensing analysis. Most of the analyses are carried out on stars, since this is expected to be a ``stress test'' maximizing the impact of undersampling.

This paper is organized as follows.
In Sec.~\ref{sec:input}, we briefly review the input data.
Section~\ref {sec:noise_corr} computes the noise power spectra of the output images for both uncorrelated (white) and correlated ($1/f$) input noise. A series of the quantitative analyses of the coadded simulated input images is presented in Sec.~\ref{sec:analysis}: we
\newcounter{objectives}
\begin{list}{\arabic{objectives}.\ }{\usecounter{objectives}}
\item use injected stars to test the output normalization, astrometry, size, and shape of the final coadd PSF after propagation through {\sc Imcom} (Sec. \ref{subsec:inj_mom});
\item measure the properties (i.e., shape and size) of the stars based on the truth positions of the input stars (Sec. \ref{subsec:psf_mom});
\item measure correlation functions of the ellipticities of both injected and simulated stars at scales overlapping the range likely to be included in the Roman weak lensing analysis (Sec. \ref{subsec:corrs});
\item measure correlations of the 4th moments of the injected stars (Sec. \ref{ss:4thmom}); and
\item estimate the noise-induced additive bias on star ellipticities using the injected stars and the Monte Carlo noise realizations, and compare this to analytic expectations (Sec. \ref{subsec:noise_bias}).
\end{list}
Section~\ref{sec:wide} presents a synthetic wide band image (averaging all 4 bands, smoothed to a common PSF) and the ellipticity correlations of simulated stars. We consider directions for future work in Sec.~\ref{sec:discussion}. The appendices cover analytic models for the 2D noise power spectra (Appendix~\ref{app:out-noise}) and the analytic theory of additive noise bias (Appendix~\ref{app:noise-additive}).

\section{Simulated Data}
\label{sec:input}

This study utilizes the simulated data products from Paper~I, which covers 0.64 square degrees in four filters: Y106, J129, H158, and F184 from shortest to longest wavelength. 
The sky inputs are from the Large Synoptic Survey Telescope Dark Energy Science Collaboration Data Challenge 2 (LSST DESC DC2; \citealt{2019ApJS..245...26K, 2021arXiv210104855L, 2021ApJS..253...31L, 2022OJAp....5E...1K}).
Table~\paperitablelayers\ in Paper~I illustrates various input images that have been fed into the {\sc Imcom} algorithm, which can be classified as follows. 
\begin{itemize}
    \item {\em Sky image} (layer names: \textbf{SCI}, \textbf{truth}): The images were generated through the {\slshape Roman} image simulations pipeline (\citealt{2021MNRAS.501.2044T, 2022arXiv220906829T}) utilizing the truth catalogs of stars and galaxies from the LSST DESC CosmoDC2 simulations \citep{2019ApJS..245...26K, 2022OJAp....5E...1K}. The layer labeled \textbf{SCI} includes saturation and simple detector physics models such as read noise and dark current, simulated in {\sc GalSim}, while the layer labeled \textbf{truth} refers to noiseless images; star and galaxy profiles are drawn at designated locations. 
    \item {\em Point sources} (layer names: \textbf{gsstar14}, \textbf{cstar14}): These are grids of sources located at HEALPix resolution 14 ({\tt nside}=$2^{14}$) pixels \citep{2005ApJ...622..759G}. Each source is a $\delta$-function convolved with a simulated {\slshape Roman} PSF; the PSFs are simulated in \cite{2022arXiv220906829T} through {\sc GalSim} using flat spectral energy distribution (SED). \textbf{gsstar14} refers to images where the sources were drawn using {\sc GalSim}, while \textbf{cstar14} refers to images where they were drawn with internal routines in the coaddition code. We call them ``injected stars'' throughout this paper. 
    \item {\em Noise images} (layer names: \textbf{whitenoise1}, \textbf{1fnoise2}): We generate realizations of white noise (\textbf{whitenoise1}) and $1/f$ noise (\textbf{1fnoise2}). The white noise is uncorrelated between pixels and has unit variance. The $1/f$ noise is scale-invariant in the time stream and is then re-formatted into a 2-dimensional image according to the pixel readout order \citep{2020JATIS...6d6001M}. It has unit variance per logarithmic range in frequency.
\end{itemize}

\begin{figure*}
    \centering
    \includegraphics[width=2.25in]{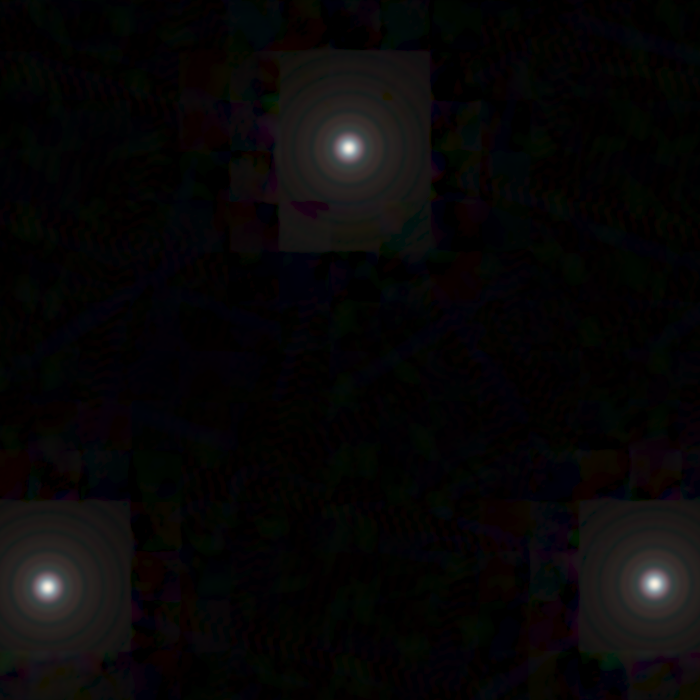}~
    \includegraphics[width=2.25in]{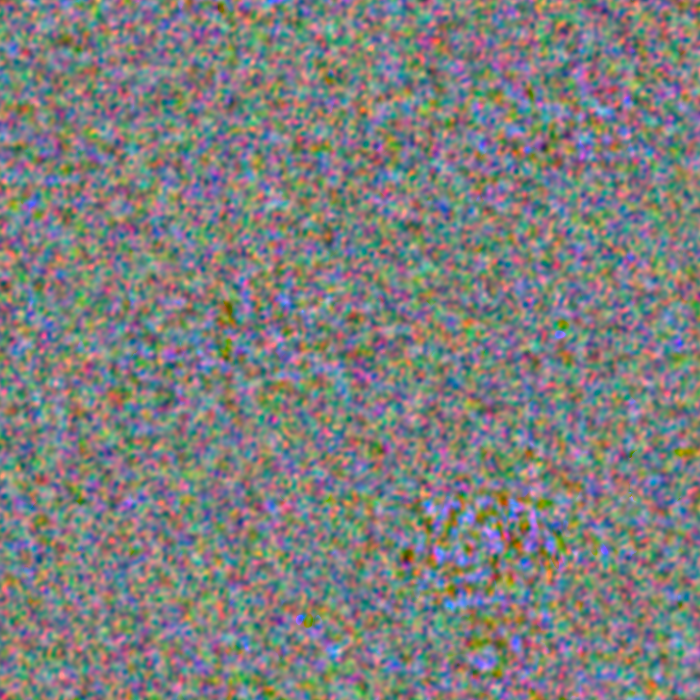}~
    \includegraphics[width=2.25in]{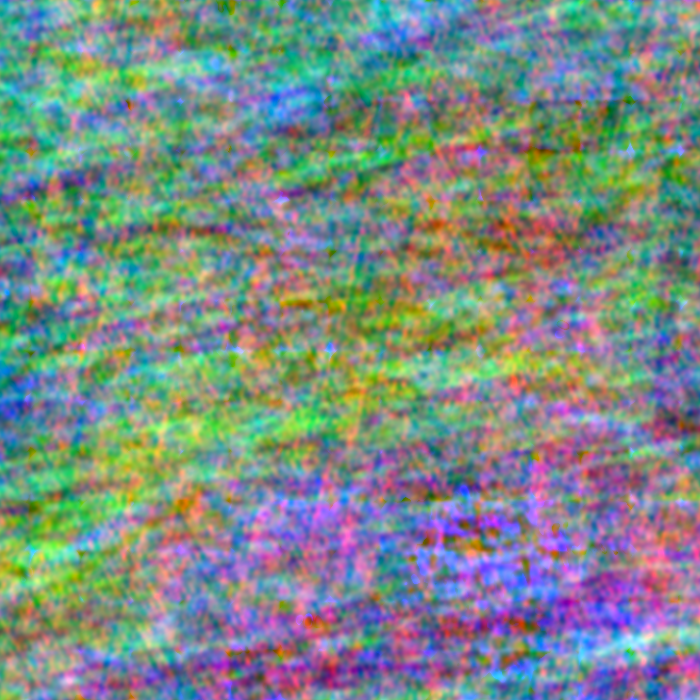}
    \caption{\label{fig:noise_images} Coadded injected {\sc GalSim} stars (left), white noise (center) and $1/f$ noise (right) realizations, displayed as 3-color F184 (red)/J129 (green)/Y106 (blue) combinations. Each image shows a $700\times 700$ output pixel ($17.5\times 17.5$ arcsec) region of the coadded images, from the \textbf{gsstar14}, \textbf{whitenoise1}, and \textbf{1fnoise2} layers, respectively. The color scale is a fourth-root stretch (0 to 0.2 input flux per input pixel) in the injected star image (left); and for the noise realizations it is linear, spanning $\pm 1.25$ (center) or $\pm 4$ (right) in input units. Note that the input white noise layer leads to output noise correlated on the scale of the input pixels, whereas the $1/f$ noise layer shows the characteristic striping at each of the 2 input rolls. These rolls are at different angles in each filter, hence the color pattern. The region shown is $270\le x<970$, $301\le y<1001$ of block (2,30).}
\end{figure*}

Examples of the injected stars and noise realizations are shown in Fig.~\ref{fig:noise_images}. And examples of a variety of objects identified in the co-added sky images are shown in Fig.~\paperifiginterestingobjects\ of Paper~I.

Throughout this paper, we use the quantity called ``fidelity'' as the basic quality indicator of the {\sc Imcom} output, and many statistics are reported as a function of the fidelity. This defines how well the output PSF recovered the target PSF we specified. Mathematically, fidelity depends on the difference between the output and target PSF,
\begin{equation}
    L_{\alpha}({\bmath s}) = \sum_{i=0}^{n-1} T_{\alpha i} G_{i}({\bmath R}_\alpha - {\bmath r}_i + {\bmath s}) - \Gamma_{\alpha}({\bmath s}),
\end{equation}
where the index $\alpha$ indicates an output pixel; $i$ indicates an input pixel; $G_i$ is the PSF in the $i$th input pixel; and ${\bmath R}_\alpha$ and ${\bmath r}_i$ are the output and input pixel positions. The fidelity is defined as the square norm of the output PSF residual scaled to the square norm of the target PSF, and re-written as an inverse logarithmic measure:
\begin{equation}
    {\rm Fidelity} \equiv -10 \log_{10} \frac{U_{\alpha}}C
    \equiv -10 \log_{10} \frac{\int [{L}_{\alpha}(\mathbf{s})]^2 d^2\mathbf{s}}{\int [{\Gamma}(\mathbf{s})]^2 d^2\mathbf{s}}.
\end{equation}
The fidelity is usually --- but not always --- better (larger) if there are more exposures. The fidelity map in our simulation footprint for each bandpass we simulated can be found in Fig.~\paperifigfidelitymap\ of Paper~I.

\section{Measurement of noise correlations}
\label{sec:noise_corr}

We first investigate the noise fields, giving an overview of noise effects (Sec.~\ref{ss:noise-overview}), and then describing the 2D (Sec.~\ref{ss:noise-2D}) and 1D azimuthally averaged (Sec.~\ref{ss:noise-1D}) noise power spectra.

\subsection{Overview}
\label{ss:noise-overview}

Noise is a significant source of error for the precision necessary to observe weak gravitational lensing of galaxies. The impact of noise on the measurement of a shear signal can be parameterized into two types of systematic observational errors: additive and multiplicative biases \citep[e.g.][]{2006MNRAS.368.1323H}. If the true signal is given by $\gamma_{\rm true}$, noise (or other) biases give a $\gamma_{\rm meas}$ of
\begin{equation}
\gamma_{\rm meas} = (1+m)\gamma_{\rm true}+c.
\end{equation}
The additive bias (also called ``spurious shear'') is given by the constant \textit{c}, and manifests as a shear signal that is present even when the true population of galaxies is unlensed.  The factor of \textit{m} gives the multiplicative bias (also called ``calibration error''). Multiplicative bias occurs when a real lensing signal is detected, but the measurement is larger or smaller than the true signal by some factor \citep{2001MNRAS.325.1065B, 2001A&A...366..717E, 2003MNRAS.343..459H}. The next-generation weak lensing surveys have stringent requirements for both types of errors \citep{2008A&A...484...67P, 2013MNRAS.429..661M, 2013MNRAS.431.3103C, 2018arXiv180901669T, 2020A&A...635A.139E, 2021MNRAS.501.2044T}.

To quantify the impact that noise might have on observations and determine whether these requirements will be met, we measure and analyze noise correlations in the coadded images from two types of input noise fields: white noise and $1/f$ noise (see Sec.~\paperisecnoise\ of Paper~I). Real noise has both of these components, as well as some smaller additional terms \citep{2015PASP..127.1144R}. We focus on the noise power spectrum in the output images, since both the lowest-order noise-induced bias and the variance of the measured shapes are proportional to second moments of the noise (i.e., depend on (S/N)$^{-2}$; \citealt{2002AJ....123..583B}) and are therefore captured by the power spectrum. We then compare this power spectrum to the expected output noise power spectrum if we ignored sampling issues. Appendix~\ref{app:noise-additive} describes in detail the impact of anisotropic noise (such as $1/f$ noise and non-ideal white noise) on ellipticity measurements. 

Different properties of a galaxy are affected by noise at different wave numbers. If a large scale uniform (zero wavenumber) noise offset is present in an image, the error in this flux would impact the estimated photometric flux from the galaxy, but not the position or shape.
An astrometric measurement will, however, be affected by an overall tilt (gradient of the noise, or in Fourier space the noise weighted by one factor of wavenumber); this would bias the centroid towards one direction or another. The shape of a galaxy will only be biased if the second derivative of the noise field is biased towards a preferred direction. In Fourier space, the ellipticity error thus comes from the noise weighted by two powers of the wave number (the impact of noise on shapes is discussed more in Sec.~\ref{subsec:noise_bias}). 
Photometry, astrometry, and shape measurements are thus dependent on higher and higher spatial frequencies respective to one another. This is demonstrated quantitatively in Appendix~\ref{app:noise-additive}, including the factor of wave vector squared ($u^2-v^2$ for $g_1$ and $2uv$ for $g_2$) appearing in the noise-induced error in the shapes.

\subsection{2D noise power spectra}
\label{ss:noise-2D}

We take the convention that the power spectrum $P(u,v)$ of a field is given by the 2D Fourier Transform of the correlation function $\xi(\Delta x,\Delta y)$:
\begin{equation}\label{eq:P2D}
P(u,v) = \int \xi(\Delta x, \Delta y) {\rm e}^{-2 \pi {\rm i}(u\Delta x + v\Delta y)} \,{\rm d}\Delta x\,{\rm d}\Delta y,
\end{equation}
where $\xi(\Delta x, \Delta y) = \langle S(x,y) S(x',y') \rangle$
is the correlation function of the noise field $S$ with pixel coordinates given by $(x,y)$ in Cartesian space and $(u,v)$ in Fourier space. In the case of discrete data, we replace the integral with a sum over pixels; we preserve the convention that $P$ has units of $S^2\times\,$area (e.g., for a field sampled at the output pixel scale $s_{\rm out}$ as considered here, we sum over pixels and multiply by $s_{\rm out}^2$).

\begin{figure*}
    \centering
    \includegraphics[width=6.5in]{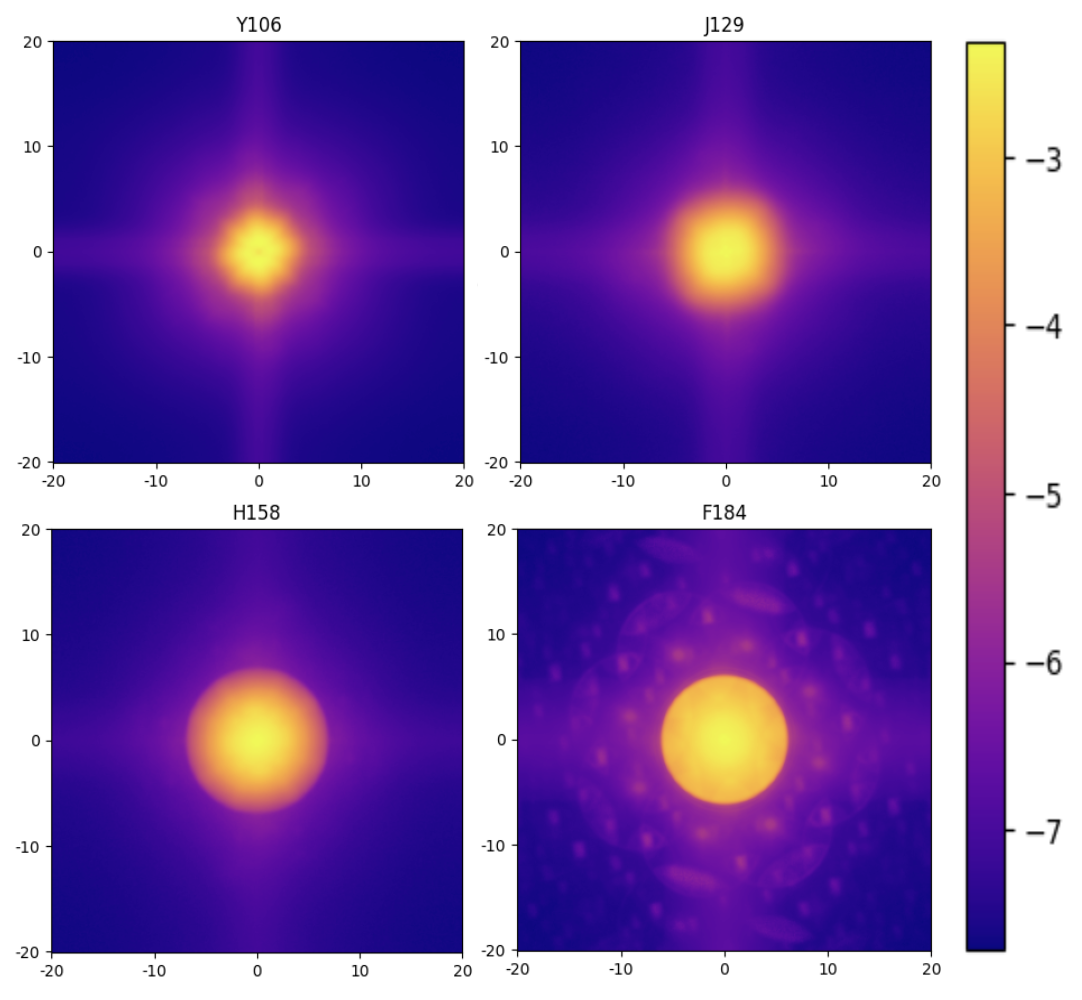}
    \caption{\label{fig:2dspectra_w} 2D averaged power spectrum of the white noise field of each band, plotted on a logarithmic color scale. The horizontal and vertical axes show wave vector components ($u$ and $v$ respectively) ranging from $-20$ to $+20$ cycles arcsec$^{-1}$. The color scale shows the power $P(u, v)$ in units of arcsec$^2$ (Eq.~\ref{eq:P2D}). The minimum and maximum values of each power spectrum are as follows: {$6.6 \times 10^{-8}$} to {$6.7 \times 10^{-3}$} for Y106; $3.4 \times 10^{-8}$ to $3.5 \times 10^{-3}$ for J129; {$5.9 \times 10^{-8}$} to {$3.2 \times 10^{-3}$} for H158; and {$2.0 \times 10^{-8}$} to {$4.8 \times 10^{-3}$} for F184.}
\end{figure*}

We accomplish this by making use of the Numpy Fast Fourier Transform (FFT) function \citep{2020Natur.585..357H}. Each noise field in each $N\times N$ pixel block (here $N=2600$) is FFTed and normalized by the size of the blocks and the size of the output pixels:
\begin{equation}
P_{2D}(u,v) = \frac{s_{\rm out}^2}{N^2}\left|
\sum_{j_x,j_y} S_{j_x,j_y} {\rm e}^{-2\pi{\rm i}s_{\rm out}(uj_x+vj_y)}
\right|^2,
\end{equation}
where $u$ and $v$ are sampled at integer multiples of $1/(Ns_{\rm out})$.
The two-dimensional power spectra that result from this are further binned into $8\times 8$ bins, resulting in power spectra that are $325 \times 325$ pixels covering the range of $u$ and $v$ from $-20$ to $+20$ cycles\,arcsec$^{-1}$ (the Nyquist limit for $s_{\rm out}=0.025$ arcsec/pixel) at spacing
\begin{equation}
\Delta u = \Delta v = \frac{1}{325s_{\rm out}} = 0.123 \,{\rm cycles\,arcsec}^{-1}.
\label{eq:dudv}
\end{equation}
For each observation filter, we then averaged together the power spectra from each of the 2304 blocks (the total $48 \times 48$ arcmin$^2$ region). The resulting average power spectrum for the full mosaic in each observing filter can be seen in Fig.~\ref{fig:2dspectra_w}.
Note that this procedure included the padding regions that appear in more than one block, so these regions are over-represented in the averaged power spectrum. Since the padding regions are not special relative to the detectors or tiling strategy, we do not think this is a major issue.

In the absence of sampling issues and if the input and output PSFs were all identical, input white noise should lead to output white noise and a constant power spectrum for the 2D image. Since we have an output PSF that is larger than the input PSF in real space (narrower in Fourier space), we expect that the high-wave number Fourier modes will have reduced power (proportional to the square of the Fourier transform of the output PSF). The examples in Figure \ref{fig:2dspectra_w} clearly show this behavior. Note that the spectra are plotted on a log scale, so that although several features are visible, any features outside of the central maxima have very little power. The circular region in the center of the 2D power spectra comes from the output PSF: the Airy disc convolved with a Gaussian. In the Y106 and J129 bands, where the Gaussian dominates the cutoff of the PSF, we see a smooth fade into uncorrelated wavenumber space. In the H158 band and especially the F184 band, the Airy disk defines the limit of the PSF, so beyond this limit the image contains only noise --- this causes {\sc Imcom} to set the background to zero, causing the hard edge of the circle.

We can confirm this by comparing measurements of the image with expectations. The band limit (maximum number of cycles per unit angle) for a diffraction-limited telescope is given by $D/\lambda$, where $D$ is the telescope diameter and $\lambda$ the wavelength of light. For the F184 band,
\begin{equation}
\frac D\lambda = \frac{2.37\,\text{m}}{1.84\,\mu\text{m}} = 1.29\times 10^6 {\rm \,cycles\, rad}^{-1}
= 6.25 {\rm \,cycles\, arcsec}^{-1}.
\end{equation}
This is indeed the radius of the circle on the F184 band image, confirming that this feature is caused by the Fourier modes outside the circular regions being beyond the target PSF band limits.

The large ``$+$ signs'' extended throughout the images correspond to the directions of the postage stamp boundaries. (Note that the input images are at various roll angles, so these features must instead be associated with the output.) The step function-like feature of the postage stamp boundaries becomes a continuous line in Fourier space, so we see these extended $+$ signs in the power spectra.

The spots in the F184 band are a small print-through of the initial pixel positions in the input images into the final output images. The first ring of 8 points is located at a radius of $\sim 9$ cycles arcsec$^{-1}$ from the center, i.e., roughly the inverse of the input pixel scale $s_{\rm in}=0.11\,$arcsec. As expected, the directions of the 8 points corresponds to the 4 grid directions of the input exposures.

\begin{figure*}
    \centering
    \includegraphics[width=6.5in]{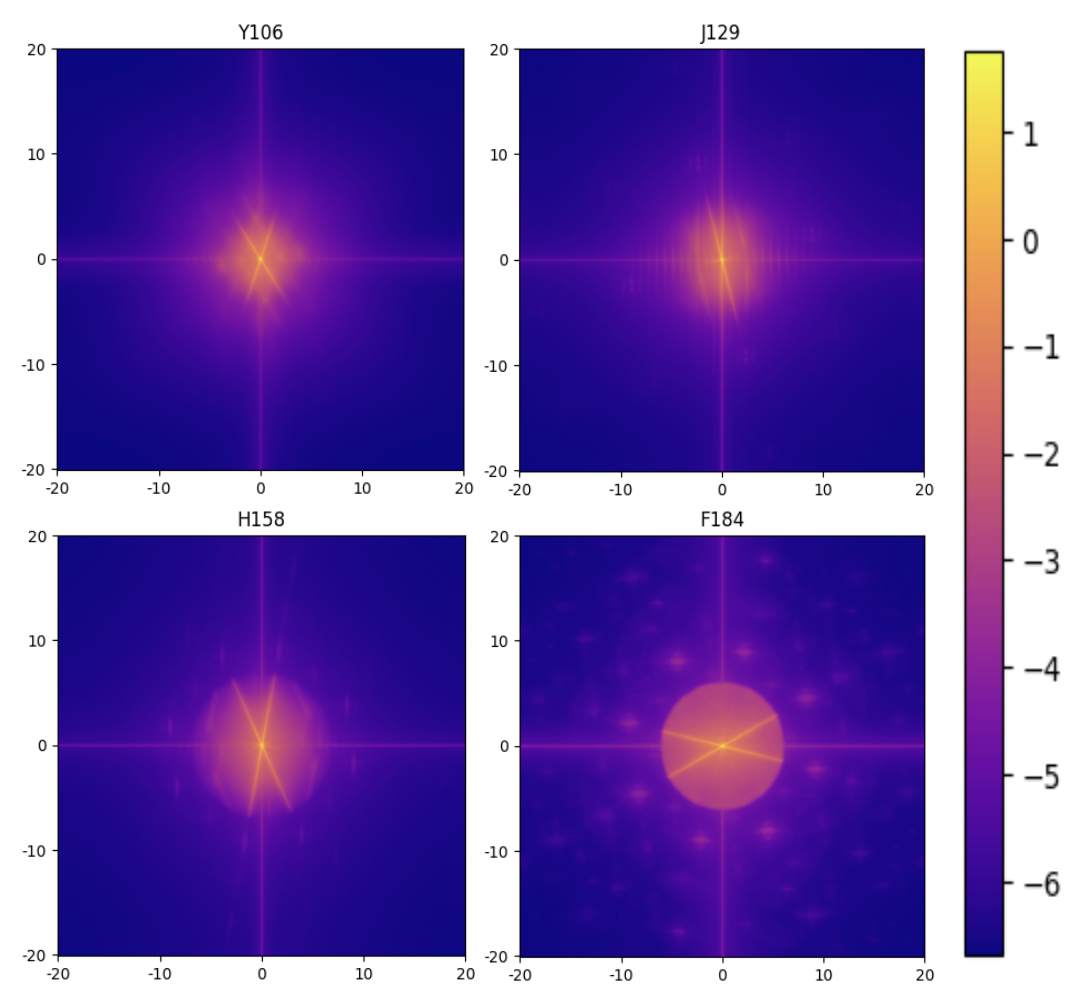}
    \caption{\label{fig:2dspectra_f}The 2D averaged power spectrum of the coadded $1/f$ noise field of each band, plotted on a logarithmic color scale. The horizontal and vertical axes show wave vector components ($u$ and $v$ respectively) ranging from $-20$ to $+20$ cycles arcsec$^{-1}$. The color scale shows the power $P(u, v)$ in units of arcsec$^2$ (Eq.~\ref{eq:P2D}). The minimum and maximum values of each power spectrum are as follows: {$1.8 \times 10^{-6}$} to {$5.4 \times 10^1$} for Y106; $8.5 \times 10^{-7}$ to {$4.6 \times 10^1$} for J129; {$6.9 \times 10^{-7}$} to {$4.8 \times 10^1$} for H158; and {$2.1 \times 10^{-7}$} to {$5.6 \times 10^1$} for F184. The X-shape, $+$ sign, spots (in H158 and F184), and vertical fringes (in J129) are discussed in the main text.}
\end{figure*}

For the $1/f$ noise power spectra in Figure \ref{fig:2dspectra_f}, we have combined two sets of images with horizontal banding at different roll angles, so the output noise image will have at least two distinct preferred directions (see the right panel of Fig.~\ref{fig:noise_images}). These features appear strongly in each panel of Figure \ref{fig:2dspectra_f} as a distinct X through the centers of the images. The roll angles are slightly different in each band in order to maximize coverage, leading to different orientations of the X's in each image. 

As in the white noise spectra, the boundary between postage stamps contained in the images creates step-like features in the noise fields, which manifest as two distinct perpendicular Fourier modes in the power spectra. However, this feature is clearly more prominent in the $1/f$ noise than in the white noise. In $1/f$ noise, we have striping across entire channels within the detector, which then crosses over postage stamp boundaries. The correlations between postage stamp boundary-caused features extend over much larger scales in real space, corresponding to sharper features in Fourier space. Thus we see the postage stamp boundary feature (the large $+$ sign) more distinctly in these images.

The $1/f$ bands also display the same reduction of power at large wave number as in the white noise, and the sharp circular boundary in H158 and especially F184 caused by the band limit of the PSFs.

In addition to features coinciding with behaviors seen in the white noise images, we see one additional pattern appear in the $1/f$ noise that is not present in white noise data. Figure~\ref{fig:2dspectra_f} shows this feature most clearly in the J129 band: alternating bright and dark vertical fringes across the center of the 2D power spectrum image. We measure the spacing between fringes to be 0.8 cycles arcsec$^{-1}$, or one cycle per 1.25 arcsec postage stamp. We therefore believe that the fringes result from steps at the postage stamp edges, with a sense that is coherent across multiple postage stamps (due to the large correlation length of the $1/f$ noise).

While these rough analytical arguments allow us to identify specific features in the output 2D power spectra with properties of the input images and {\sc Imcom} algorithm, the quantitative details --- how much power appears in each feature, and how this varies as a function of sampling going from Y106 (bluest/worst sampled) to F184 (reddest/best sampled) --- must be determined via simulations.

\subsection{Azimuthally averaged power spectra}
\label{ss:noise-1D}

In addition to the two-dimensional power spectra, we generate for each filter a set of one-dimensional azimuthally averaged power spectra. First, the two-dimensional power spectra of each block are saved from the previous step of analysis. We then calculate the mean exposure coverage in each block, and group the power spectra into one of 5 bins in mean coverage. Since the root-mean-square of noise in the output noise images shows the mean exposure coverage dependence, it is good to check how noise correlations depend on the coverage as well (see Sec.~5.3 of Paper~I for more details). These power spectra are also separated by observing filter, as exposure coverage varies significantly depending on the band in which observations are taken. 
We average together the 2D power spectra for blocks within a given mean coverage bin into a single 2D power spectrum, which is then azimuthally averaged over thin annuli, using the method from \citet{2023MNRAS.tmp..359C}. For this work we used 162 radial bins, effectively taking rings of width one pixel in the power spectrum map (Eq.~\ref{eq:dudv}) and averaging the values of the power at each circular aperture of wavenumbers. 

\begin{figure*}
    \centering
    \includegraphics[width=7in]{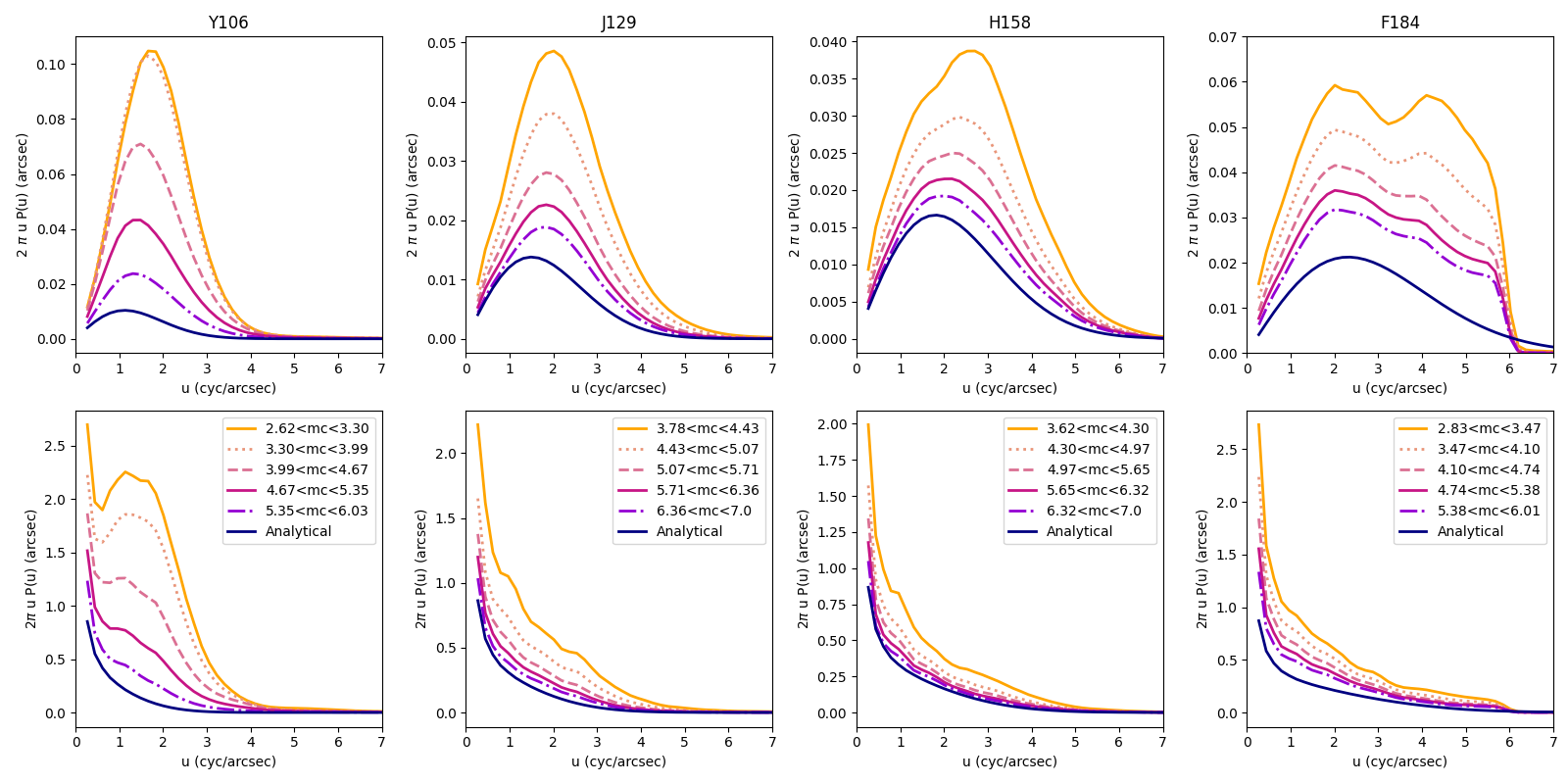}
    \caption{\label{fig:1dspectra} 
    Top row: 1D power spectra for the output white noise fields in each filter. Bottom row: 1D power spectra for the output  $1/f$ noise fields in each filter. Each filter's spectra are divided into five even-width bins of mean coverage (``mc'' in short), and plotted against the analytical expectation for noise power spectra for combining 5 exposures in the absence of sampling issues (see Appendix \ref{app:out-noise} for derivations). }
\end{figure*}

Figure~\ref{fig:1dspectra} shows the mean coverage-binned 1D power spectra for the output white and $1/f$ noise images in each observation filter.
Rather than just $P(\umag)$, we plot $2 \pi \umag P(\umag)$ so that the area under the curve corresponds to the variance in the output pixels. We additionally include in each figure a purely analytical model of the 1D power spectrum, based on $N_{\rm in}=5$ input exposures and ignoring pixelization/sampling issues. Derivations of the analytical models for the output noise power spectra can be found in Appendix \ref{app:out-noise}.

For all bands and for both types of noise, the peak power and width of the power spectrum depends consistently on the mean coverage in the image. Exposures with small mean coverage values have the highest peaks and the slowest dropoffs down to zero power at large wave number. Our analytic expectation represents an idealized case, and the behavior as compared with the spectra for the mean coverage bins shows that with decreasing mean coverage, we move farther above the ideal power spectrum. However, we note that in all of these cases the noise power spectrum is above the analytic expectation for most wave numbers.

\begin{figure*}
\centering
    \includegraphics[width=6.5in]{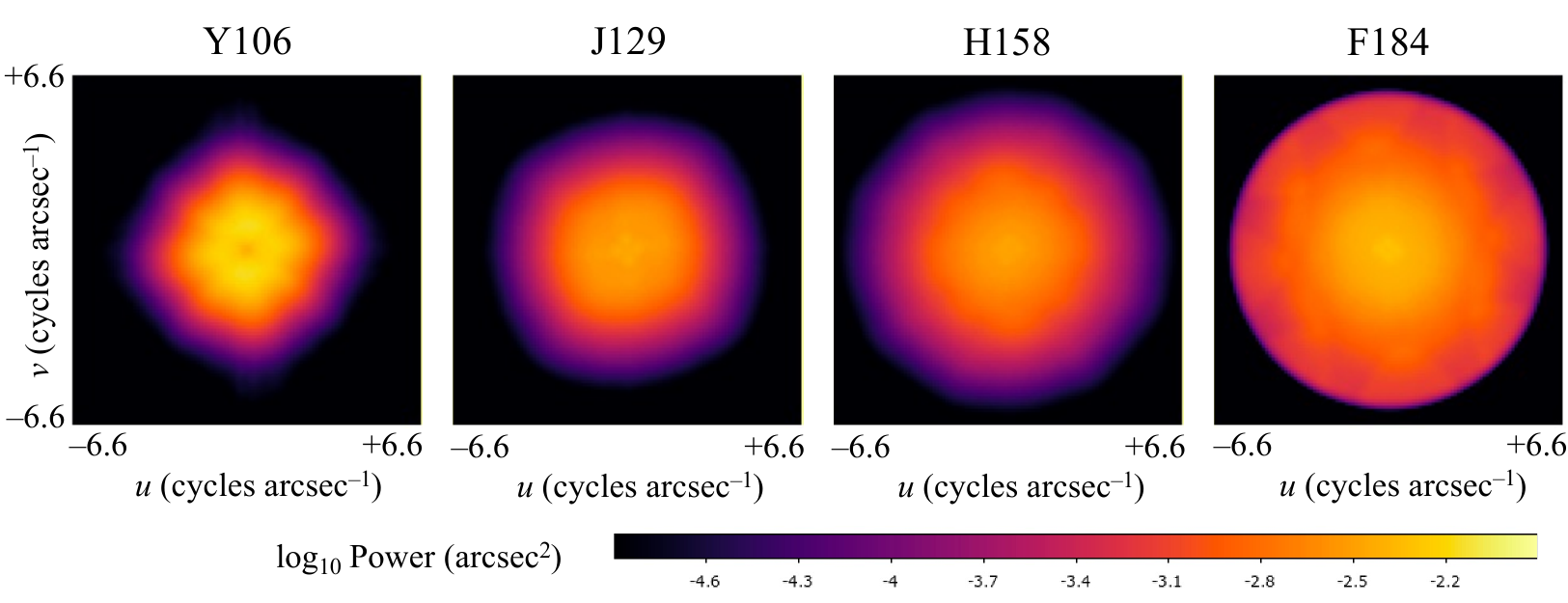}
    \caption{\label{fig:2d_wn_zoom} 
    Zoomed in image of the central features in the input white noise power spectra. By stretching the color scale we can see more clearly that the zero-wavenumber modes do not contribute the most to power in the output noise, particularly in Y106.}
\end{figure*}

In the white noise 1D power spectra, several features are evident in all bands. Each of the observation bands shows that the contributions to the power spectrum (weighted by $2\pi\upsilon$) go up, rise to some peak, and come back down to zero, qualitatively consistent with the analytic expectation. Noise increase due to aliasing often gets worse as one moves to larger wave number (the zero-mode corresponds to total flux, which is conserved in the absence of intrapixel sensitivity variation). This can result in $2\pi\upsilon P(\upsilon)$ having a steeper dependence than $\propto\upsilon$ at small $\upsilon$, as seen prominently for Y106 (the most undersampled filter) in Fig.~\ref{fig:1dspectra}. Figure~\ref{fig:2d_wn_zoom} shows the central feature in the 2D power spectrum from input white noise for each filter with the color scale significantly stretched to show this behavior.

In contrast, the $1/f$ noise spectra have a distinct peak in the center of the image, caused by the overlaps in the Fourier modes from the roll angles. The power spectrum as a function of frequency on one dimension goes like $1/f$, so for small wave numbers the power should reach very large values --- a result that is clearly visible in the spectra for all bands. The analytic estimation for the $1/f$ noise spectra in the absence of pixelization/sampling effects (Appendix~\ref{app:out-noise}) is qualitatively correct for the redder filters, but the normalization is slightly low, and it misses the ``bump'' caused by aliasing in the bluer filters (especially Y106).

Analyzing these power spectra allows us to understand the correlations between features caused by noise in the simulated images being operated on by {\sc Imcom}. In Section~\ref{subsec:noise_bias}, we provide a more detailed  discussion of the quantitative links between these power spectra and weak lensing measurement biases.

\section{Moments Analysis of the coadded images}\label{sec:analysis}

In this section, we present the quantitative analysis of the simulated coadded images of multiple layers. This includes object centroid, ellipticity and size measured by computing image moments of objects located in grids, and simulated fields with and without noise.

\subsection{Moments of the injected sources}\label{subsec:inj_mom}

Our first set of tests is conducted on the grids of simulated stars. Our main objective is to show the properties of the output PSFs and demonstrate that these properties have an expected dependency on how divergent the output PSF is from the target. The expected properties are exactly known since we chose the target PSF to be the Airy disk convolved with the Gaussian kernel.

Our simulation footprint contains 54\,597 unique simulated stars, each at the center of a HEALPix\footnote{http://healpix.sourceforge.net} \citep{2005ApJ...622..759G} resolution 14 pixel. For each injected star, we cut out a $99\times 99$ output pixel (2.475$\times 2.475$ arcsec) postage stamp from the output block containing that star (for the ``block'' definition refer to Fig.~\paperifigtilefig\ of Paper~I). We consider the first the {\sc GalSim} noiseless star layer (\textbf{gstar14}).\footnote{There is also a \textbf{cstar14} layer: this should in principle be the same except for approximations used to draw the stars. We have computed the correlation function $\xi_+(\theta)$ from the nearest-neigbor ($\approx 0.2$ arcmin) to the diagonal (1 degree) in all 4 bands, and found values ranging from consistent with zero up through $2.1\times 10^{-9}$. Given that this difference is small compared to {\slshape Roman} requirements, we present only the {\sc GalSim}-drawn stars (which are independent of any of the {\sc Imcom} routines) in the main text to avoid clutter.} We measure the $1^{\rm st}$ and $2^{\rm nd}$-order moments of the simulated sources in the output images using the {\tt galsim.hsm} module \citep{2003MNRAS.343..459H, 2005MNRAS.361.1287M}, which implements the fitting of an adaptive elliptical Gaussian to the image \citep[e.g.][]{2002AJ....123..583B}.\footnote{The {\tt galsim.hsm} module can implement PSF corrections to galaxy shapes using the method of \citet{2003MNRAS.343..459H} and \citet{2005MNRAS.361.1287M}, but this functionality is not used in this section since we are only measuring the stars.} The $1^{\rm st}$ moments (centroids) can be compared with the expected position based on the World Coordinate System (WCS) of the coadded image; the result is reported as the astrometric error. The $2^{\rm nd}$ moments (covariance of the Gaussian: a $2\times 2$ matrix ${\mathbfss M}$) are reported as three real numbers: the shear-invariant width
\begin{equation}
\label{eq:size_definition}
\sigma = \sqrt[4]{M_{xx}M_{yy}-M_{xy}^2},
\end{equation}
and the shape components
\begin{equation}
\label{eq:shape_definition}
(g_1,g_2) = \frac{(M_{xx}-M_{yy}, 2M_{xy})}
{M_{xx} + M_{yy} + 2\sqrt{M_{xx}M_{yy} - M_{xy}^2}}
\end{equation}
that are zero for a perfectly circular object and generally satisfy $g_1^2+g_2^2<1$ for a positive definite second moment matrix.
(The use of ``$g$'' for the latter indicates consistency with the conventions of \citealt{1995A&A...294..411S}.) We also report the mean fidelity $\langle -\log_{10} (U_\alpha/C)\rangle$ for the pixels in the central $20\times 20$ output pixel (i.e., $0.5\times 0.5$ arcsec) region surrounding the star. This is a measure of how well {\sc Imcom} estimates it has done at building an output image with the desired PSF in that region of the survey.

\begin{figure*}
    \centering
    \includegraphics[width=6.75in]{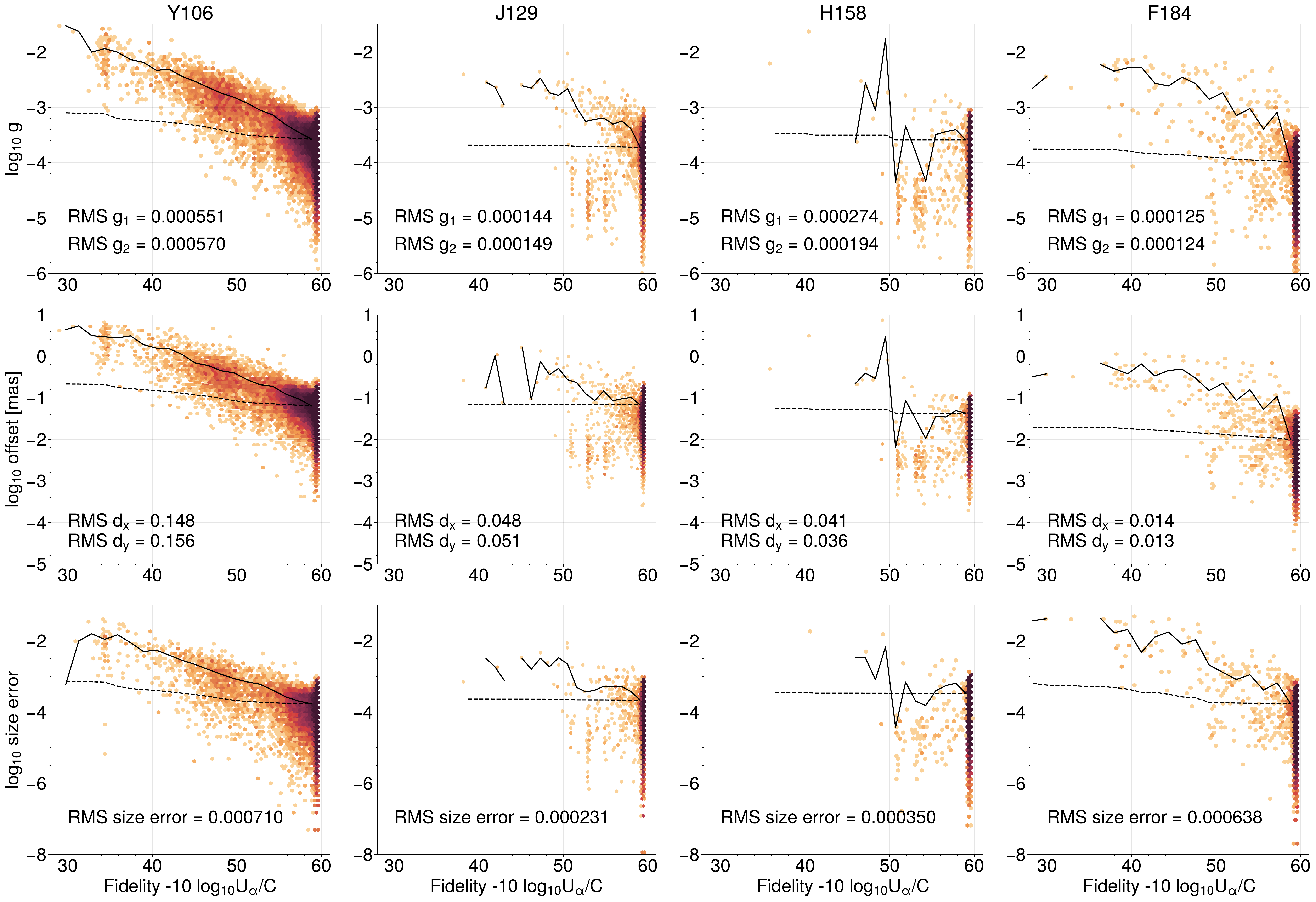}
    \caption{\label{fig:starstatfig_v2}The distributions (in an arcsinh scale) of moments of the injected stars drawn in the input images, coadded in {\sc Imcom}, and then measured by the {\tt galsim.hsm} module in each of the filters, with no noise. \textbf{\emph{Top}}: the ellipticity $g=\sqrt{g_1^2+g_2^2}$ of the injected {\sc GalSim} stars. \textbf{\emph{Middle}}: the astrometric displacement from the coordinates where the star was injected. \textbf{\emph{Bottom}}: the relative size error of the {\sc GalSim} stars and target PSF. The definition of size in this figure is Eqn.~\ref{eq:size_definition}. The horizontal axis is the output PSF fidelity (larger means a better match to the target PSF; see Sec.~\ref{sec:input}). The solid line shows the RMS ellipticity (top), RMS astrometric displacement (middle), and RMS fractional size error (bottom) in each fidelity bin. The dashed line is the same thing but is cumulative (i.e., the RMS for all stars in regions with that fidelity or better). This would be applicable to cases where we impose a mask based on the fidelity.
    }
\end{figure*}

Figure~\ref{fig:starstatfig_v2} shows measured ellipticity, astrometric error and fractional size difference of the sources drawn by {\sc GalSim}. It is expected that the anisotropy of shapes decreases as the fidelity of the output image is maximized. Most of the postage stamps (for the ``stamp'' definition refer to Fig.~\paperifigtilefig\ of Paper~I) achieved high fidelity (above 50) in all bandpasses except Y106. This is likely due to the bandpasses being the most undersampled and that the algorithm is unable to find the transformation matrix that the leakage and noise correlations are minimal (this is also shown in Fig.~\paperifigurefcum\ of Paper~I). In all bandpasses, however, the total RMS ellipticity of the coadded injected stars in each shape component is $\lesssim$ $5.7 \times 10^{-4}$ which is the PSF ellipticity requirement on angular multipole scales (32 < $\ell$ < 3200) determined by the SRD. Here we also demonstrate that the astrometric error of the sources induced by {\sc Imcom} is a small contribution to the relative error budget in the astrometric calibration ($< 1.3$ mas) defined by the SRD. Finally, the bottom panel of Fig.~\ref{fig:starstatfig_v2} shows the fractional size error of the injected stars relative to the target PSF. We cannot directly compare the RMS size error in each bandpass to the required PSF size error in the SRD ($\lesssim$ $3.6 \times 10^{-4}$)\footnote{The requirement on relative size error $\sigma_{\rm T}/T$ (where $T = M_{xx} + M_{yy} = 2\sigma^2/(1-g^2) \approx 2\sigma^2$) can be propagated to $\sigma_{\rm \sigma}/\sigma \sim\sigma_{T}/2T$.} because our RMS errors are computed at individual points, whereas the size error requirement is computed at scales $\ell<3200$ (or smoothed at a scale of $\sim 4$ arcmin, i.e., where there are $3200^2$ pixels on the full celestial sphere). We see that the size errors are largest in Y106 and F184. If we bin the stars into pixels, we see that the size error declines to $(1.58,0.59,1.01,2.09)\times 10^{-4}$ in Y106/J129/H158/F184 respectively in $1\times 1$ arcmin pixels; and then it declines to $(0.46,0.39,0.70,1.46)\times 10^{-4}$ in $4\times 4$ arcmin pixels. Thus we see that the image combination step is using only a small fraction $(1.46/3.6)^2 = 16\%$ of the overall PSF size error budget in a root-sum-square sense.

\begin{figure*}
\includegraphics[width=6.4in]{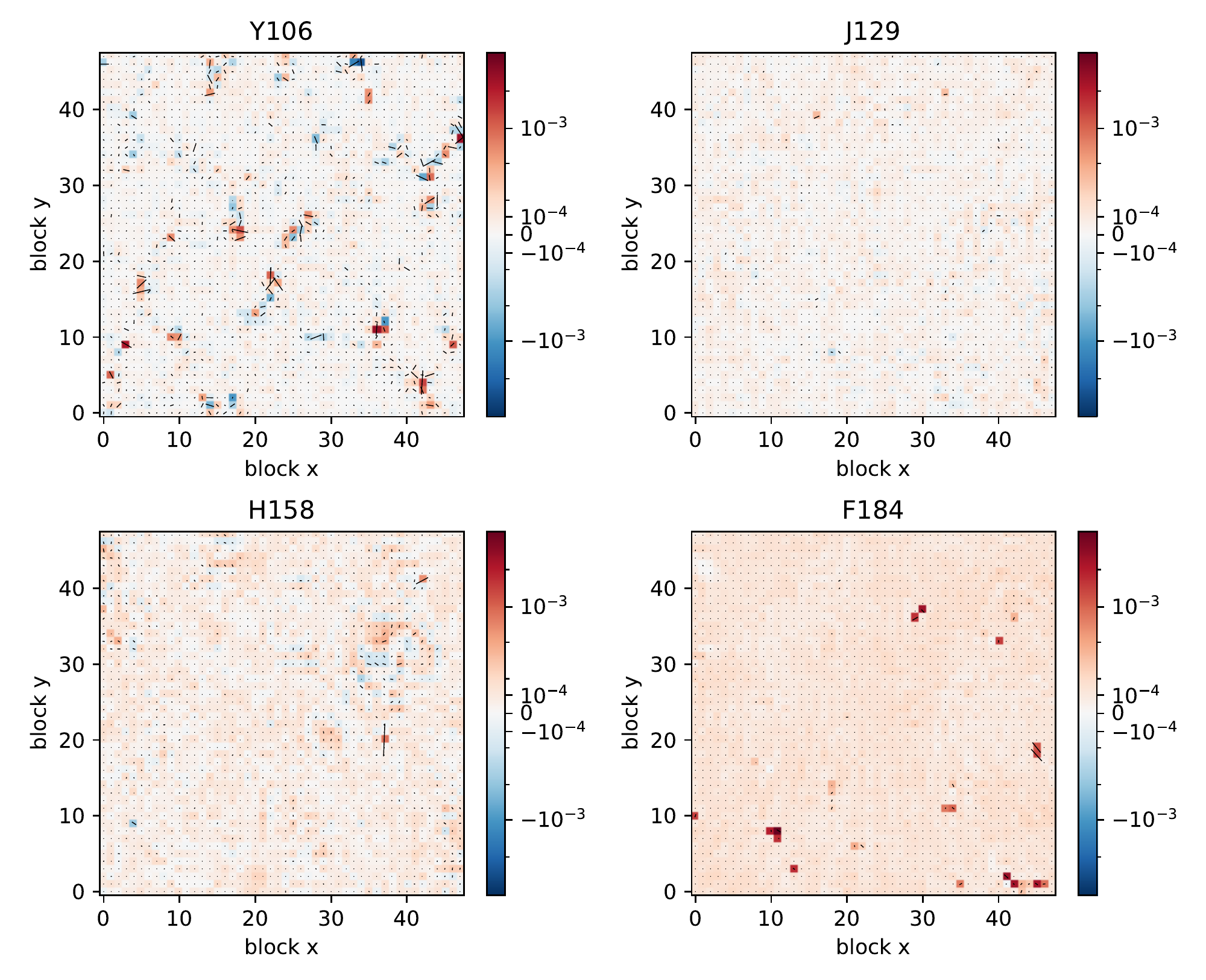}
\caption{\label{fig:etmap}A spatial map of the size error (color scale) shape (whiskers) of the injected stars. Each panel shows the $48\times 48$ arcmin footprint of the simulation in one of the 4 filters. Stars are averaged into pixels (1 arcmin$^2$ each). The color scale shows size error $(\sigma_{\rm star}/\sigma_{\rm target}-1)$ on an arcsinh scale, and the whiskers are scaled so that a length of 1 pixel corresponds to $g=5\times 10^{-4}$. This is presented on a square grid, however we have checked that the main features are also present if gridded in HEALPix and so they are not gridding artifacts.}
\end{figure*}

The spatial distribution of the second moment errors, averaged in 1 arcmin$^2$ pixels, is shown in Fig.~\ref{fig:etmap}. Most of the area has ellipticity errors at the $\lesssim 10^{-4}$ level; one sees a few outliers, with corresponding errors in the size. The large outliers seen in Y106 and F184 correspond to low-coverage regions (see Fig.~\paperifigcoveragemap\ of Paper~I). What matters most for weak lensing purposes are the 2-point statistical features of these maps; the correlation function will be investigated in Sec.~\ref{subsec:corrs}.

We note that these quantities are also measured on the sources drawn by our internal ``croutines'' and we find consistent results as {\sc GalSim} sources.

\subsection{Measurement of stars in the simulated images}\label{subsec:psf_mom}

For precise measurements of galaxy properties, of particular interest is the propagation of the PSF through coaddition algorithms. In principle, {\sc Imcom} tries to remove any anisotropy or inhomogeneity from the coadded PSF and coadd stars should result in being round. However, since the leakage (square norm of the difference of output and target PSF) is not exactly zero there is some residual ellipticity of the coadded stars even in the absence of noise. Although stars in the coadded images are not utilized directly for modeling the PSF (with {\sc Imcom}, the PSF determination step must be performed on single-epoch images prior to coaddition\footnote{{\sc Imcom} requires PSF models for each exposure to calculate correlations between PSFs.}), the coadded stars are an important test of PSF propagation and any ellipticity that we see at the end will have implications for galaxy measurement. 

Stars are identified in the coadded sky images based on the truth locations of these stars. We use the simple simulated sky images for {\slshape Roman} simulated in \citet{2022arXiv220906829T}, which utilized the {LSST DESC DC2} simulations \citep{2021ApJS..253...31L}. Its stellar catalog was based on \textit{Galfast} \citep{2008ApJ...673..864J, 2018ascl.soft10001J} simulations. We cut out postage stamps around the truth locations and measure the star properties using the {\tt galsim.hsm} module (which allows the flux, centroid, and second moments to float). The truth catalog contains locations, magnitudes estimated in \emph{Roman} bandpasses from object flux based on objects' SEDs, and whether the star is a candidate for PSF modeling. The criteria for being a candidate star defined in \citet{2022arXiv220906829T} are that:
\begin{itemize}
    \item no pixel in the simulated star stamp is saturated in at least one exposure of the star (saturation was chosen at $10^5$ e in the images with \emph{simple} detector models); and
    \item the star has a detection signal-to-noise (S/N)$_{\rm det}$ above 50, as defined by (S/N)$_{\rm det}$ = $0.015 \times (\rm total\ flux)$, where 0.015 was the typical background inverse noise level. (The observed signal-to-noise ratio including source Poisson noise is slightly lower.)
\end{itemize}
The S/N level was chosen here to be permissive to allow more restrictive selections later. We used this criteria for selecting clean stellar samples to conduct our measurement. We additionally select stars whose magnitudes are fainter than 18 in all bandpasses since we found that some single-exposure images contained saturated stars. (In the real survey, some information might be obtained from saturated stars because the detectors are read non-destructively and the early, non-saturated reads can be used; however, it is possible that the PSF would be different since one does not average over telescope motion during the full exposure, and in any case the present simulations do not include the early reads.)

\begin{figure*}
    \centering
    \includegraphics[width=6.5in]{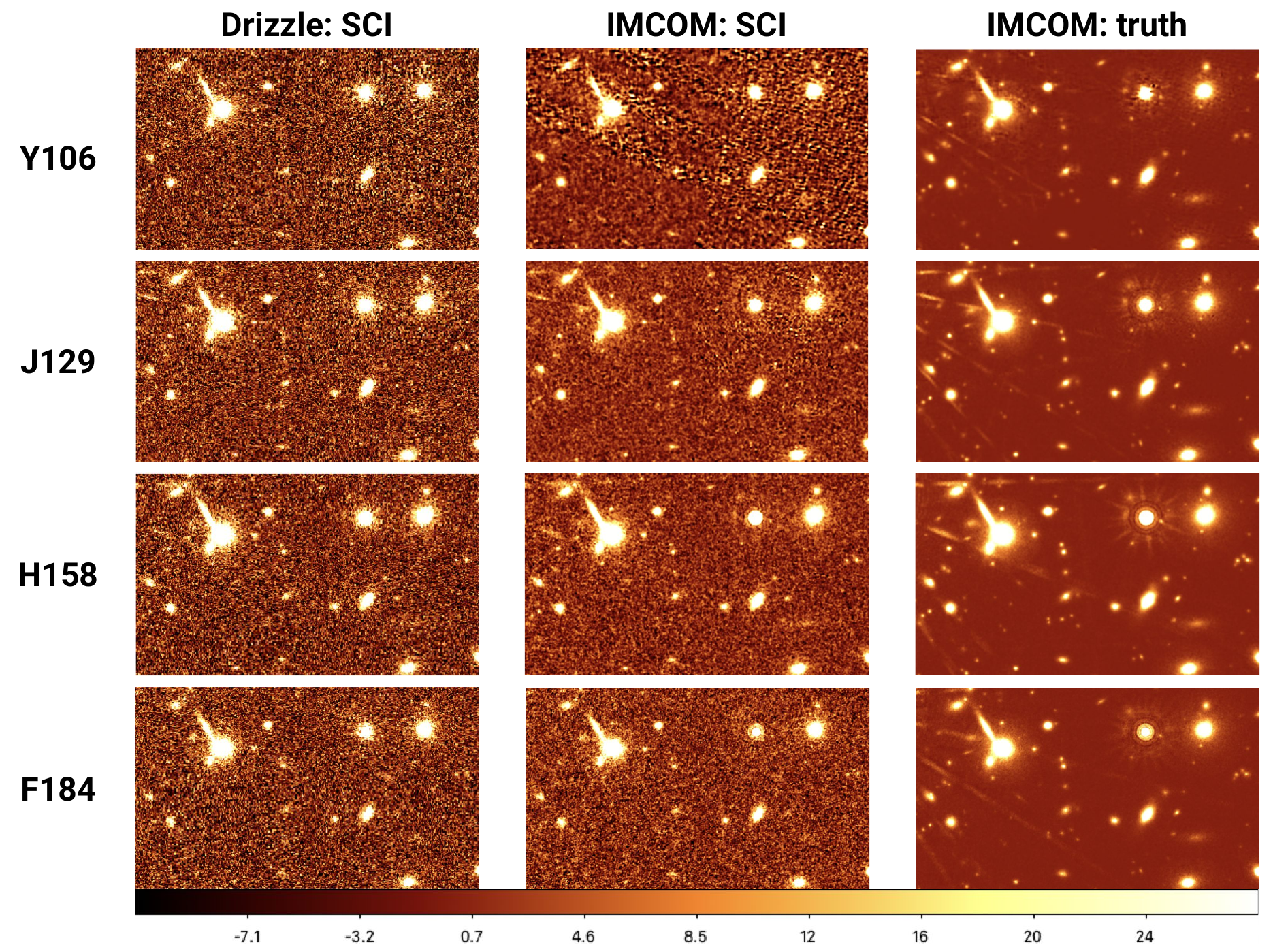}
    \caption{\label{fig:star_fig} Coadded images of \textbf{SCI} and \textbf{truth} layers for a selected region for Y106, J129, H158 and F184 bandpasses. \textbf{\emph{Left}}: {\sc Drizzle} co-added \textbf{SCI} image, \textbf{\emph{Middle}}: {\sc Imcom} co-added \textbf{SCI} image. \textbf{\emph{Right}}: {\sc Imcom} co-added \textbf{truth} image. The flux in {\sc Drizzle} images are scaled by (0.11/0.0575)$^2$ to account for the difference in the output pixel size and normalized to show the same color range. The area that is shown here is $32.7 \times 19.2$ arcsec centered around RA$=53.006\,$deg and Dec$=-40.027$. Note that this is near the center of our simulated output region.
    }
\end{figure*}

With these selections, we consider 2623 (Y106), 2674 (J129), 2588 (H158), 2647 (F184) stars in our simulation footprint. For each star in each of the {\sc Imcom} images (\textbf{SCI} and \textbf{truth}), we cut out $99 \times 99$ output pixels postage stamps resulting in a stamp 2.475 arcsec on each side, whereas we extract $43 \times 43$ pixels around the truth centroid of a star in {\sc Drizzle} images resulting in a stamp 2.473 arcsec on each side. Although the cutouts from {\sc Drizzle} coadded ``SCI'' images (\textbf{{\sc Drizzle}: SCI}) are directly comparable with those of {\sc Imcom} coadded ``SCI'' images (\textbf{{\sc Imcom}: SCI}), the measurements of stars on ``truth'' images are not. While we cut out postage stamps from {\sc Imcom} coadded ``truth'' images (\textbf{{\sc Imcom}: truth}), {\sc Drizzle} coadded images of the ``truth'' layer were not simulated by \citet{2022arXiv220906829T} at the time of this project.
Examples of {\sc Imcom} and {\sc Drizzle} images are shown in Fig.~\ref{fig:star_fig}.

\begin{figure*}
    \centering
    \includegraphics[width=\textwidth]{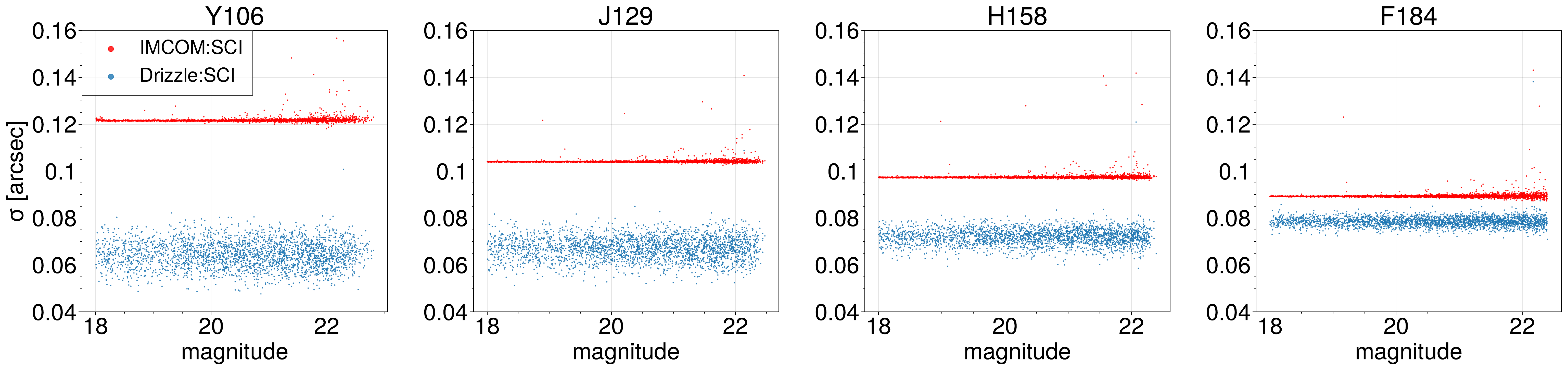}
    \caption{\label{fig:sizemag} Size-magnitude diagram for the selected stars (mag$>18$) in our simulated {\sc Imcom} and {\sc Drizzle} images. The size of the stars is measured with the adaptive moments method \citep{2002AJ....123..583B}. Note that {\sc Imcom} produces a larger output PSF, but it is much more uniform (see main text).
    }
\end{figure*}

Compared to the measurement of injected stars (Sec. \ref{subsec:inj_mom}), the measurement on the cutout of stars is expected to contain complications that could directly affect the shape and size of an object. These complications include background and simulated-detector noise, blending, and chromaticity (since the injected sources are drawn with the same SED passed to {\sc Imcom}, but the simulated stars are drawn with their ``true'' SEDs). The simulated stars therefore test more of the sources of systematic biases in weak lensing surveys. Some of these effects can be seen in Figure~\ref{fig:sizemag} as outliers from the stellar locus in each bandpass. Fig.~\ref{fig:sizemag} additionally displays stellar locus of the same stars in {\sc Drizzle} images, which is located at a smaller size than {\sc Imcom} stars. This is expected because {\sc Drizzle} does not smear the PSF. However, there is a large dispersion in sizes especially in the bluer, more undersampled bands. The {\sc Drizzle} algorithm is designed to preserve total flux in its ``raining down'' procedure \citep{2002PASP..114..144F}, but the spatial spread in the output image will depend on the specific locations of the pixels. Even though the output PSF from {\sc Imcom} is larger than {\sc Drizzle} (especially in Y106), it is designed to be uniform: it must be large enough so that that output resolution could be achieved even in a part of the image where the pixels interlace each other differently (see Paper~I, Fig.~\paperifigyresfig). The {\sc Imcom} output PSF is still much smaller than ground-based PSF and the narrowness of the locus will help with star-galaxy separation in the real mission. 

\begin{figure*}
    \centering
    \includegraphics[width=0.8\textwidth]{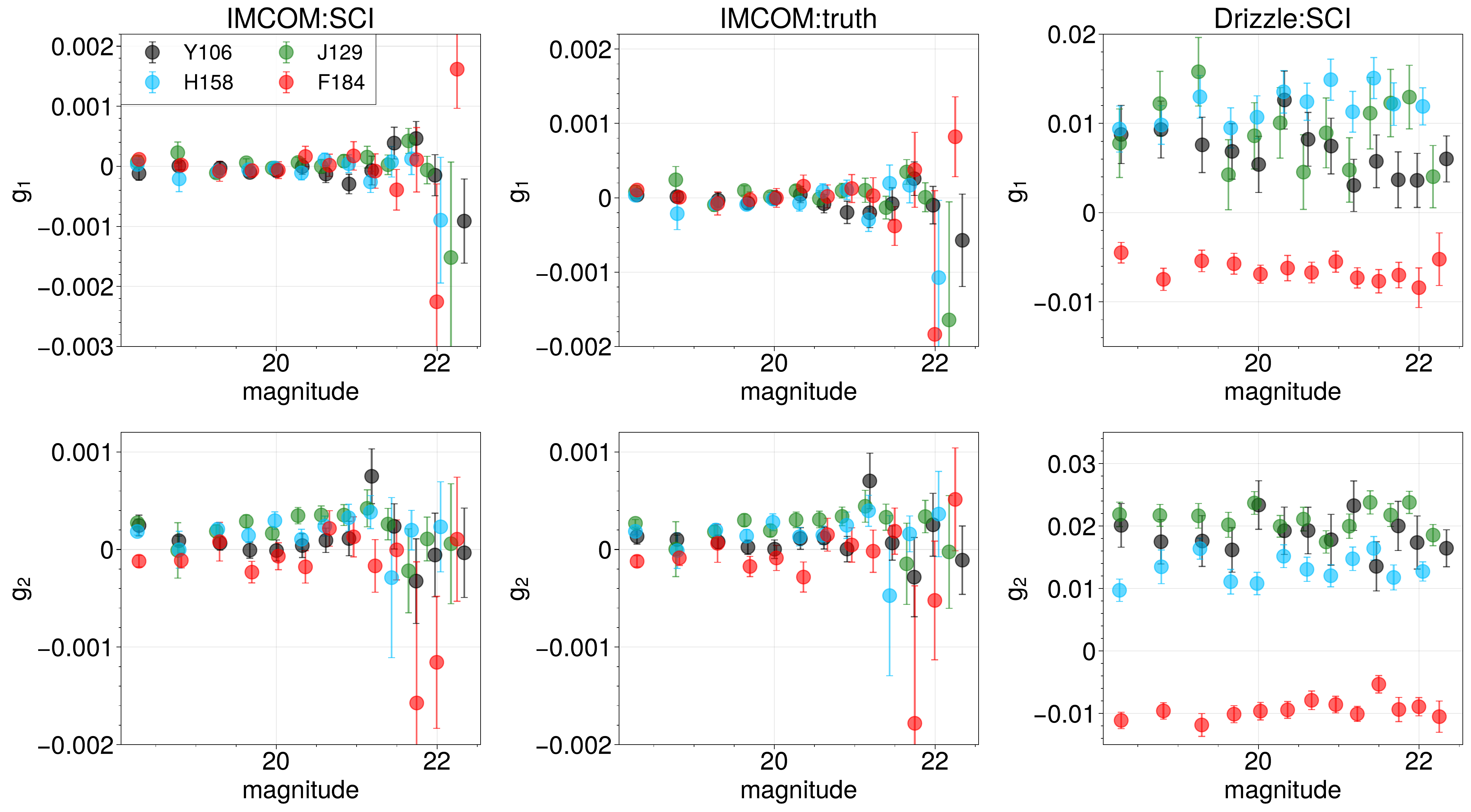}
    \caption{\label{fig:psfstar} The mean ellipticity ($g_1$, $g_2$) of PSF candidate stars as a function of star magnitude of corresponding bandpass is shown. Four colored data points in a panel show the values from the stars found in Y106, J129, H158, F184 images. These stars were cut out from mosaics of coadded images using the coordinates from truth star catalog. The magnitude is also taken from the truth catalog. The ellipicity is measured with adaptive moments method in {\sc GalSim} and the error bar is a standard error of the mean. \textbf{\emph{Top}}: the ellipticiity ($g_1$) from the cutouts of three sets of images ({\sc Imcom} and {\sc Drizzle} coadded \textbf{SCI} images, {\sc Imcom} coadded \textbf{truth} images. \textbf{\emph{Bottom}}: $g_2$ for the same set of stars. We use the {\sc Drizzle} images produced in \citealt{2022arXiv220906829T}.
    }
\end{figure*}

Figure~\ref{fig:psfstar} shows the measured ellipticity of stars found in simulated images as a function of object magnitude. The mean ellipticity components $g_1$ and $g_2$ in {\sc Imcom} are at the level of a $\lesssim $few$\times 10^{-4}$ for all bandpasses. By comparison, the measured shapes on {\sc Drizzle} images are consistently different from zero by an amount of order $\sim 10^{-2}$ (\textbf{{\sc Drizzle}: SCI}). It shows that the {\sc Drizzle} process is not able to average out the shape of PSFs, regardless of the fact that output pixel size is larger in {\sc Drizzle} images (0.0575 arcsec/pixel) than {\sc Imcom} (0.025 arcsec/pixel) and pixel size does have an effect on the measured PSF ellipticity (Fig.~E1 of \citealt{2021MNRAS.501.2044T}).

\begin{figure}
    \centering
    \includegraphics[width=\columnwidth]{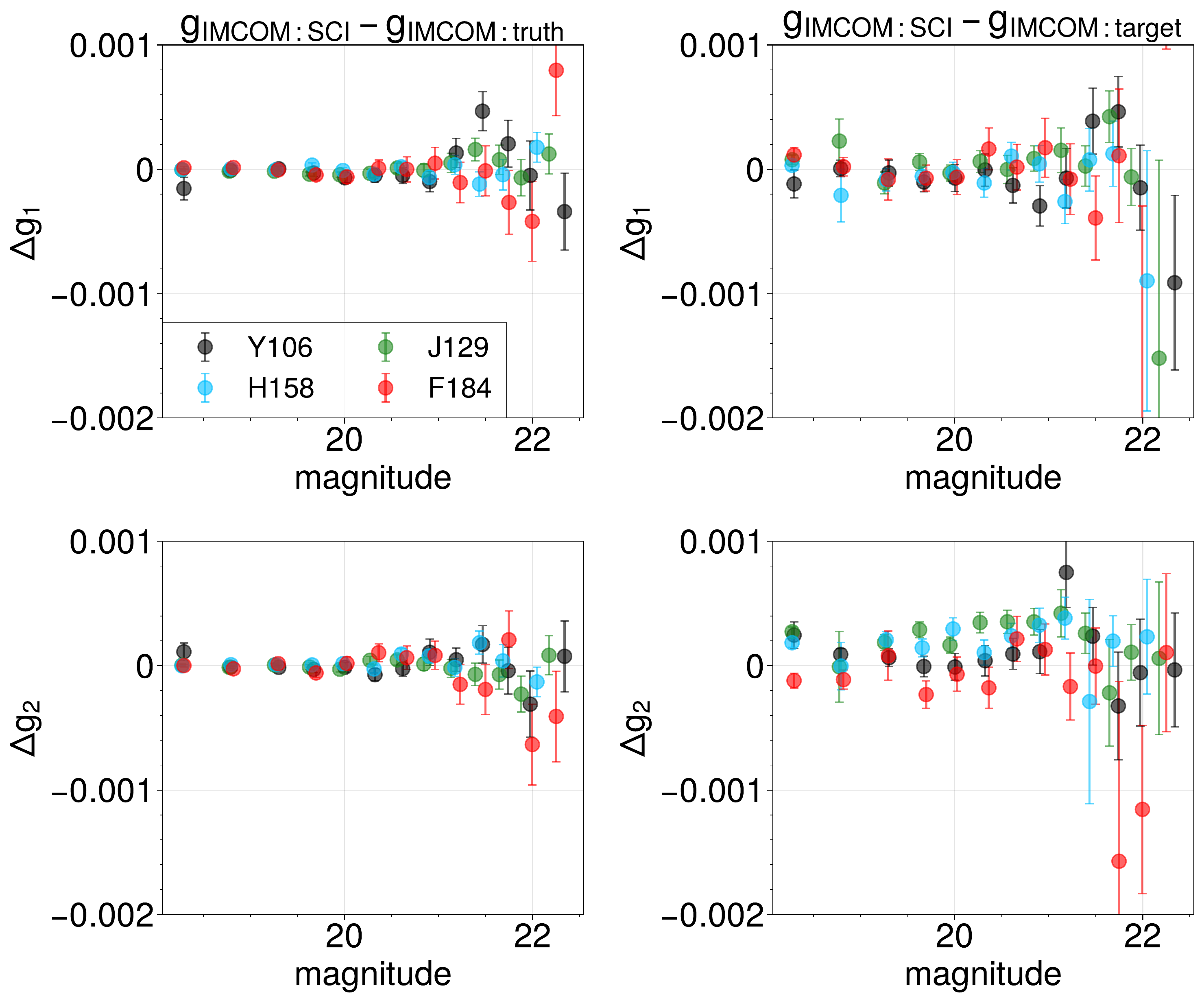}
    \caption{\label{fig:psfstar_residual} The residual ellipticity ($g_1$: \textbf{\emph{Top}}, $g_2$: \textbf{\emph{Bottom}}) of the stars extracted from several layers of simulated images is presented. \textbf{\emph{Left}}: the difference in the shapes from {\sc Imcom} coadded \textbf{SCI} and \textbf{truth} images. This shows the effect of simple noise models on the ellipticity as \textbf{SCI} images include them and \textbf{truth} images do not. \textbf{\emph{Right}}: the difference in the shapes from the {\sc Imcom} coadded \textbf{SCI} images and target PSF we chose in our {\sc Imcom} simulations. This shows the cumulative effect of noise in sky images and output PSF fidelity.
    }
\end{figure}

Figure~\ref{fig:psfstar_residual} shows the residual ellipticity measured on the cutouts of stars from ``SCI'' and ``truth'' images, and the ellipticity compared to the target PSF for {\sc Imcom} (which in this case is only different from zero due to numerical precision). The difference between ``SCI'' and ``truth'' shows the residual shape introduced by the simple detector physics noise model (which includes dark current, saturation, and read and Poisson noise). Here, the average residual ellipticity is $\lesssim$ $2\times10^{-4}$ for bright stars and $\lesssim$ $8\times10^{-4}$ for faint stars, verifying that {\sc Imcom} shows a tolerance to simple noise models on star shapes. We also investigated whether achieving the target fidelity can affect star shapes of different magnitudes. The right column of Fig. \ref{fig:psfstar_residual} that shows the cumulative effect of noise and fidelity indicates that for $g_1$ the residual behavior is consistent with the simple noise model case (top left panel of the same figure) whereas $\Delta g_2$ shows a non-zero residual ellipticity over a range of magnitude. Although this is likely caused by the output PSF that did not exactly match the target, the leakage into star shapes is less than the PSF ellipticity error requirement.

\begin{figure*}
    \centering
    \includegraphics[width=\textwidth]{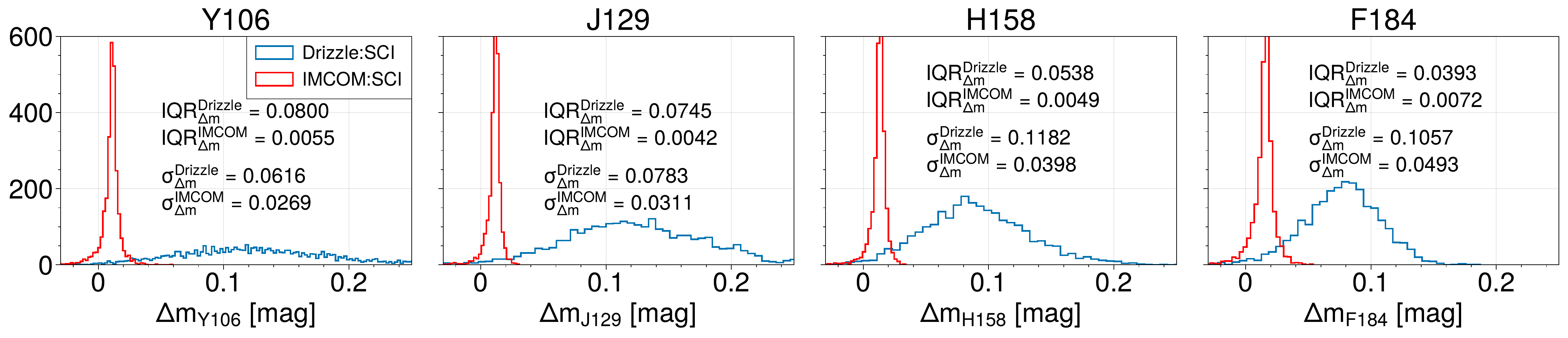}
    \caption{\label{fig:star_mag} The deviation of measured magnitudes of the stars in {\sc Drizzle} and {\sc Imcom} images from the magnitudes from the truth catalog for all the bandpasses is shown. The magnitude of the stars is estimated from the flux measurement using {\tt galsim.hsm}. Since the flux units in {\sc GalSim} are photons/cm$^2$/s, we multiply the flux measured on {\sc Imcom} star cutouts by (0.025/0.11)$^2$ to obtain flux from surface brightness. There is no need for this scaling for {\sc Drizzle} cutouts because {\sc Drizzle} operates on fluxes rather than surface brightnesses. The number of photons is then converted to the AB magnitude through Eq.~(\ref{eq:magab}). The texts in each panel display the interquartile range (IQR) to indicate the spread that is unaffected by outliers, and the standard deviation of the sample.
    }
\end{figure*}

Additionally, we have estimated the magnitudes of the stars from the total image intensity for best-fit elliptical Gaussian in {\sc GalSim}. These magnitudes are estimated on the stars of both {\sc Imcom} and {\sc Drizzle} ``SCI'' images. Figure~\ref{fig:star_mag} shows the relative error of the measured magnitude compared to the truth for each filter. It is estimated from the photon flux in the following way. The measured number of photons ($N_{\rm star}$) can be used to measure the magnitudes of the stars, using
\begin{equation}\label{eq:magab}
    m_{AB} = -2.5 \log_{10}\frac{N_{\rm star}}{N_{\rm zero~point}} - m_{\rm correct},
\end{equation}
where $N_{\rm zero~point}$ is the number of photons that {\slshape Roman} would observe for a 0 AB magnitude source \citep{1983ApJ...266..713O} and $m_{\rm correct}$ is an aperture correction term to account for the fluxes in the wings of stars, because the diffraction wings of the PSF are not captured by the best-fit Gaussian; thus the HSM fluxes are less than the ``total'' (integrated to $\infty$) fluxes. We compute the zero point based on the model throughput curve of {\slshape Roman} that was used in the simulation:
\begin{equation}
    N_{\rm zero~point} = \int (3.631 \times 10^{-23}\,{\rm W\,m^{-2}\,Hz^{-1}}) A_{\rm eff}(\lambda) t_{\rm obs} \frac{{\rm d}\nu}{h\nu},
\end{equation}
where $A_{\rm eff}$ is the effective collecting area of the telescope for a given bandpass, $t_{\rm obs}=139.8\,$s is the exposure time, and $h\nu$ is the energy of photons (where $h$ is Planck's constant). The effective collecting area for a given bandpass can be integrated over the bandpass\footnote{\url{https://roman.gsfc.nasa.gov/science/WFI_technical.html} Our effective area is calculated using the values from early design phase to be internally consistent with the area used in \cite{2022arXiv220906829T} simulations.}, and $\int (A_{\rm eff}/\lambda) {\rm d}\lambda$ = 0.5915, 0.6051, 0.5978, 0.3929 m$^2$ for Y106, J129, H158, and F184 respectively. We then correct for $m_{\rm correct}$ by measuring the flux and magnitude on the target PSF for each bandpass; the corrections are 0.0977, 0.1471, 0.2180, 0.3018 mag for Y106, J129, H158, and F184 respectively. 

Since we specify the target PSF in {\sc Imcom} to be the Airy disk convolved with the Gaussian kernel, the adaptive moments method captures the fluxes from these stars quite well. Their RMS residual magnitudes are 28, 32, 41, 51 mmag in Y/J/H/F band whereas the SRD states relative photometric calibration in each filter to be better than 10 mmag. We should note that this measurement includes neighboring fluxes from nearby sources (which are {\em not} included in the SRD requirement) and stars with low fidelity (outliers in Fig.~\ref{fig:sizemag}). A visual inspection of outliers from the size-magnitude diagram in Y106 showed that most are due to leaking flux from neighboring objects (not necessarily fully blended). For the adaptive moment magnitudes measured on {\sc Drizzle} images, on the other hand, as the dispersion in the difference suggests, the coadded PSF obtained through {\sc Drizzle} is not uniform across our footprint and the magnitude measurement is not reliable in the strongly undersampled case (Y106). (We note that {\sc Drizzle} preserves {\em total flux} by construction; but the fraction of the flux captured by the adaptive moment method depends on the PSF.) Even for the weakly undersampled cases, the mean of the residual magnitude is one order of magnitude worse than the {\sc Imcom} case.  

In this section, we have only explored the moments on single-band images; the moments measured on multi-band (synthetic wide-band) images will be analyzed in Sec.~\ref{sec:wide}.

\subsection{Correlation functions of stars in various outputs}\label{subsec:corrs}

We have so far looked at one-point statistics of observed properties of stars in the coadded images of various input layers. In the end, however, it is crucial to verify that systematic biases related to undersampling and the image reconstruction process do not contaminate the cosmological signal (e.g., Fig.~12 of \cite{2022arXiv220906829T} for their estimated $\xi_{\pm}$ over 20 deg$^2$ simulated sky). We approach this by measuring two-point statistics, especially shape-shape correlations of these stars. The calculations of two-point correlation functions were performed using the publicly available \textsc{TreeCorr} package \citep{2004MNRAS.352..338J}.

\begin{figure*}
    \centering
    \includegraphics[width=7in]{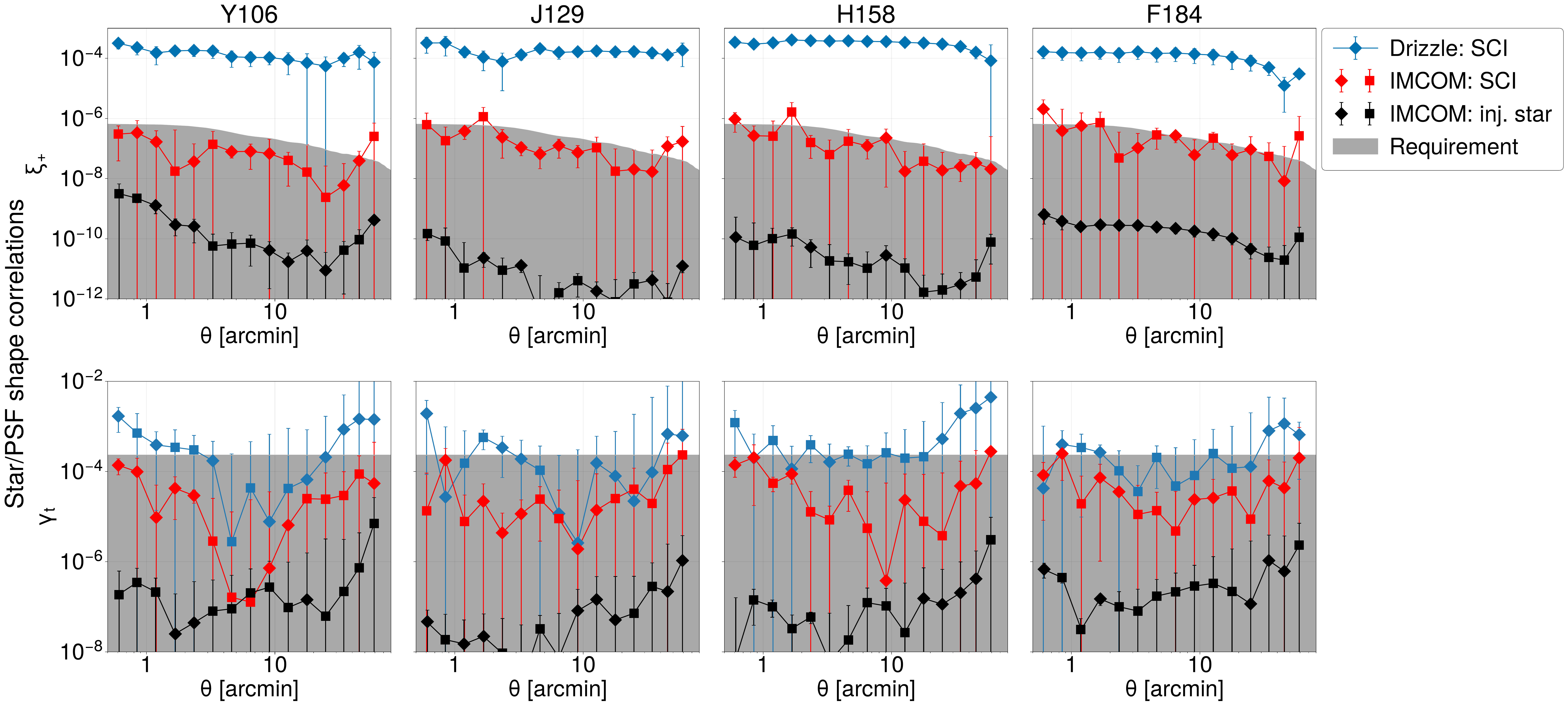}
    \caption{\label{fig:shear-shear}\textbf{\emph{Top}}: Auto-correlation ($\xi_{+}$) of observed star ellipticity from various coadded images. This includes {\sc Drizzle} and {\sc Imcom} images with simple detector noise models (see Sec.~2 for detailed explanation), and {\sc GalSim}-drawn injected stars from {\sc Imcom} images. The ellipticity was measured with adaptive moments. Dark grey shaded region shows the required $\xi_{+}$ signal approximated with the requirements on additive shear errors from SRD. Each panel displays the signals corresponding to each bandpass, and ``diamond'' marker is the positive signal while ``square'' is the negative signal but taken absolute value.
    \textbf{\emph{Bottom}}: Cross-correlation of the sky coordinates and the observed ellipticity of the same sources. Readers may refer to Fig.~\ref{fig:etmap} for the residual ellipticity around the positions where there are fewer dither positions.
    }
\end{figure*}

In Fig.~\ref{fig:shear-shear}, we measure shape-shape correlations ($\xi_{+}$) of observed stars in {\sc Imcom} coadded injected stars and ``SCI'' layers, and in {\sc Drizzle} coadded ``SCI'' image. We also show a simple comparison of these star shape correlations to {\slshape Roman} PSF mitigation requirements. Here we compute the approximate requirement on $\xi_{+}$ from the requirements on additive errors in SRD in each angular multipole moment bin. We rewrite the Hankel transform of angular power spectra $C_{\ell}$ as a Riemann sum \citep{2022PASP..134a4001G}: 
\begin{eqnarray}
    \xi_{\rm +} \!\!\!\! &=& \!\!\!\!
    \int_{0}^{\infty} \frac{\ell \,{\rm d}\ell}{2\pi}\, J_{0}(\ell \theta) (C_{EE}(\ell) + C_{BB}(\ell))
    \nonumber \\
    &\approx& 
    \!\!\!\! \sum_{\ell ~\rm bins} 2\gamma^{2} \langle J_{0}(\ell\theta) \rangle,
\end{eqnarray}
where $J_{0}$ is the Bessel function of first kind. It is clear here that the shape correlations of {\sc Imcom} coadded stars are at a statistically acceptable level compared to the requirement. The most idealized case is shown as the correlations of injected stars, drawn with exactly the same input PSF given to {\sc Imcom}: here any deviations of the measured shapes from isotropy and homogeneity can be attributed to the algorithm itself. We confirm that the level of correlations is two orders of magnitude smaller than the estimated systematic requirement. As can be seen in Fig.~\ref{fig:shear-shear}, disparities exist between $\xi_{+}$ on \textbf{{\sc Imcom}: SCI} and \textbf{{\sc Imcom}: injected stars} since the stars drawn in the noiseless injected source layer are isolated and the PSFs are made with a flat SED, while the stars drawn in \citet{2022arXiv220906829T} are fully chromatic. We should note that we do not expect the measurement using {\sc Drizzle} stars to be consistent with zero since {\sc Drizzle} does not attempt to smooth the output PSF to be isotropic.

It is worth mentioning that the $\xi_{\rm +}$ presented here is the contamination solely from star or PSF shapes --- error contributions from residual PSF shape/size errors and shape measurement are not included. We consider the shapes of the injected stars as $e_{\rm PSF}$ and those of stars in sky image as $e_{\rm star}$. Typically we characterize the additive shear systematics as the following in terms of observed shear, PSF modeling errors and noise:
\begin{equation}\label{eqn:shear_bias_model}
    \gamma^{\rm obs} = \gamma^{\rm true} + \delta e^{\rm sys}_{\rm PSF} + \delta e_{\rm noise};
\end{equation}
here $\delta e^{\rm sys}_{\rm PSF}$ (additive shear systematic biases due to PSF modeling) can be decomposed into
\begin{equation}
\label{eqn:additive_components}
    \delta e^{\rm sys}_{\rm PSF} = \alpha e_{\rm PSF} + \beta(e_{\rm star} - e_{\rm PSF}) + \eta (e_{\rm star} \frac{T_{\rm star} - T_{\rm PSF}}{T_{\rm star}})
\end{equation}
following \citet{2009A&A...500..647P} and \citet{2016MNRAS.460.2245J}. Our test in this section only quantifies the contribution to $e_{\rm PSF}$ from residuals in the image combination process, but not the coefficients $\alpha$, $\beta$, and $\eta$. This will be investigated in a future paper. We will, however, quantify the contribution to the observed shear due to additive noise biases ($\delta e_{\rm noise}$) in Sec.~\ref{subsec:noise_bias}. 

Fig.~\ref{fig:shear-shear} additionally demonstrates the tangential shear around the positions of simulated (in the ``SCI'' image) stars and the injected stars. While we expect these signals to be zero given that we have not put large scale structure in the position catalog and they seem to be below the additive shear error requirement (only in $2.0 < \log_{10}\ell < 2.5$ due to the relevant scale of galaxy-galaxy lensing), $\gamma_{\rm t}$ signal around injected stars might be underestimated because the angular distance between sources is discrete.

\subsection{Correlations of fourth moments}
\label{ss:4thmom}

In addition to the PSF second moments, \cite{2022arXiv221203257Z} demonstrated that the correlation function of the spin-2 components of the PSF higher moments (e.g., fourth moment) can cause shear additive bias to cosmic shear correlation function, contingent on the shear estimation method used in the analysis. We measure the standardized higher moments defined in \cite{2023MNRAS.520.2328Z},
\begin{equation}
\label{eq:moment_define}
    M_{pq} = \frac{\int \mathrm{d}x \, \mathrm{d}y \, u^p \, v^q \, \omega(x,y)
    \, I(x,y)}{\int \mathrm{d}x \, \mathrm{d}y \, \omega(x,y) \, I(x,y) }.
\end{equation}
Here the $(u,v)$ are transformed coordinates from the image coordinate $(x,y)$, such that the second moment shapes in Eq.~(\ref{eq:shape_definition}) vanish, and the second moment size in Eq.~(\ref{eq:size_definition}) is normalized to 1; $p$ and $q$ are the integer moment indices. For fourth moments, $p+q=4$. Here $\omega(x,y) = {\rm e}^{-{\bmath x}\cdot {\mathbfss M}^{-1}{\bmath x}/2}$ is the adaptive weight function given by HSM \citep{2003MNRAS.343..459H}, and $I(x,y)$ is the image of the PSF. The complex spin-2 fourth moment is defined as
\begin{equation}
\label{eq:spin-2-fourth-moment}
    M^{\rm (4)}_{\rm PSF} = M_{40} - M_{04} + 2{\rm i}(M_{31}+M_{13}).
\end{equation}

\begin{figure*}
    \centering
    \includegraphics[width=2.0\columnwidth]{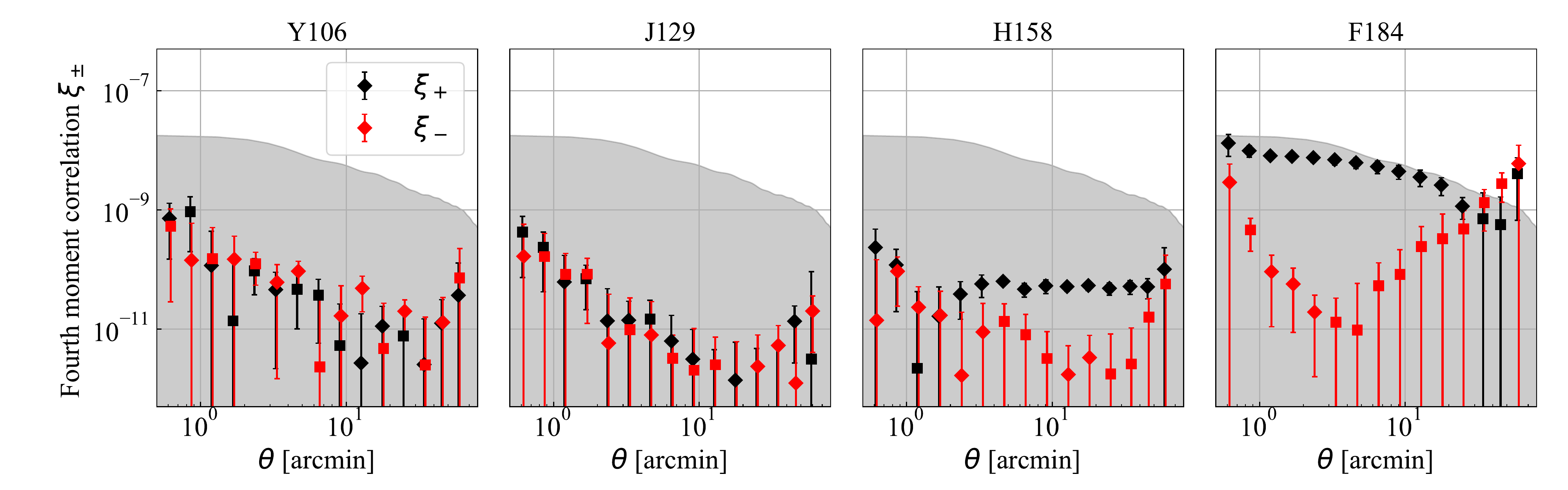}
    \caption{Auto-correlation ($\xi_{\pm}$) of the fourth moment spin-2 components ($M^{\rm (4)}_{\rm PSF}$), measured by the injected stars drawn from the {\sc GalSim} routine. Each panel displays the signals corresponding to each bandpass, and ``diamond'' marker is the positive signal while ``square'' is the negative signal but taken absolute value. The grey-shaded region shows the requirement on the fourth moment $\xi_+$, computed from the SRD and the assumption that the fourth moment leakage will be $6$-times larger than that of second moments. The PSF fourth moments correlation functions are safely within the requirement for Y106, J129, and H158 band, and are on par with the requirement for the F184 band. }
    \label{fig:fourth_correlation}
\end{figure*}

In Figure~\ref{fig:fourth_correlation}, we show the correlation function $\xi_\pm$ of the PSF fourth moments, measured on the coadded image of the injected stars drawn from the {\sc GalSim} routine. This is presented with an approximated requirement for the fourth moments correlation function from the SRD, by assuming the fourth moment leakage factor $\alpha^{\rm (2)}$ is 6 times larger than the second moment leakage factor ($\alpha$ in Eq.~\ref{eqn:additive_components}). This decision is supported by the results of \cite{2022arXiv221203257Z}. The fourth moment correlation function of the injected stars are safely within the requirement for the Y106, J129, and H158 band, showing that the higher moments of the anisotropy of the {\sc Imcom} PSF will not cause significant additive bias in these bands. However, the correlation functions in F184 are only slightly below the requirement for the most angular range and are above the requirement (though not by a statistically significant amount) for the largest scales. This indicates that the PSF fourth moments could potentially contaminate the real cosmological signal in F184. In fact, the fourth moments correlation function is $\sim 3$ order-of-magnitude larger than the second moment correlation in terms of signal-to-requirement fraction, mainly because {\sc Imcom} PSFs have extremely low large-scale coherence in the second moment shapes, which is not necessarily the same for the fourth moments.

This result suggests that the PSF fourth moments should be actively monitored in the {\slshape Roman} HLIS cosmological analysis. Further work is however required for investigating the leakage of PSF fourth moments into the galaxy shape for the {\slshape Roman} HLIS, by cross-correlating the galaxy shape with the star moments (characterization of the $\alpha, \beta, \eta$ coefficients in Eq.~\ref{eqn:additive_components}). A larger-area simulation will also be valuable here for better understanding the leakage at scales larger than 60 arcmin.

\subsection{Estimation of additive noise bias}\label{subsec:noise_bias}

A concern with measuring shapes on a coadded image is that noise correlations can result in a bias in the shape measurement even if the PSF has been made round and is exactly known \citep{2000ApJ...537..555K, 2002AJ....123..583B, 2004MNRAS.353..529H, 2012MNRAS.425.1951R, 2012MNRAS.424.2757M, 2015MNRAS.450.2963M, 2016MNRAS.457.3522G, 2017A&A...599A..73H, 2018MNRAS.479.4971O}. For example, if there are noise correlations in the $x$-direction, then there are different centroid uncertainties in the $x$ and $y$ directions; this propagates into different biases in object moments $\langle x^2\rangle$ and $\langle y^2\rangle$, and hence a bias in the ellipticity $g_1$. There are in fact a number of such terms, each resulting from the fact that the ellipticity is a non-linear function of the data, and leading overall to a bias proportional to $\nu^{-2}$ where $\nu$ is the detection significance (see, e.g., the detailed discussion in \citealt{2002AJ....123..583B}, \S8.2). This results in the {\em additive noise bias}; the aforementioned leading terms result from the second derivative of the measured shape with respect to the input image, and so we refer to it as the ``second-order'' additive noise bias. (There are third-order and higher terms as well.) A more detailed treatment will be presented in a future paper; here our aim is a first look at the second-order noise bias of the ellipticities of injected stars.

One may construct a Monte Carlo estimate of the additive bias as follows:
\begin{eqnarray}
\Delta g \!\!\!\! &=& \!\!\!\!
\frac12 \sum_{\alpha\beta} \frac{\partial^2 g}{\partial I_\alpha \,\partial I_\beta} {\rm Cov}(I_\alpha,I_\beta)
\nonumber \\
&\approx& 
\!\!\!\! \left\langle A \left[ \frac{g(I+\epsilon N) + g(I-\epsilon N)}2
 - g(I) \right] \right\rangle,
\label{eq:NoiseMC}
\end{eqnarray}
where $I$ is the image of the object; $N$ is a noise image; $A$ and $\epsilon$ are scaling factors; and the $\langle\rangle$ denotes an average over the Monte Carlo noise realizations. If the shape measurement of the object is performed on a coadded image, then $I$ should be the coadded image and $N$ should be a coadded noise realization (and thus includes any correlations imprinted by the coaddition procedure).

If the object image is normalized to have unit flux, and $N$ is normalized to be a noise image with unit input variance per input pixel, then the proper scaling is
\begin{equation}
A\epsilon^2|_{\rm white~noise} = \frac{1}{\nu_{\rm SE}^2 \Omega_{\rm psf}/s_{\rm in}^2},
\end{equation}
where $\nu_{\rm SE}$ is the single-epoch signal-to-noise ratio and $\Omega_{\rm psf}/s_{\rm in}^2$ is the input PSF effective area in pixels (shown in Table~\paperitableoutpsf\ of Paper~I). Then Eq.~(\ref{eq:NoiseMC}) returns an estimate of the second-order additive noise bias. In principle, the use of multiple Monte Carlo realizations of the noise should allow us to reduce the uncertainty in Eq.~(\ref{eq:NoiseMC}). But even a single realization is unbiased and can be used in a correlation function code such as {\sc TreeCorr} to estimate the additive noise bias correlation function.

There may also be noise bias from the input $1/f$ noise. Typically the $1/f$ noise is specified by a ``knee frequency'' $f_{\rm knee}$ (with the pixels time-ordered, so units of Hz) where the $1/f$ noise power spectrum is equal to the white noise component power spectrum. For readout, the $4096\times 4096$ pixel arrays used by Roman split into 32 channels of 4096 rows and 128 columns each. The pixels in each row are read out with a cadence of 5 $\mu$s; after all the pixels in one row are read out, we move to the next row (so the time to read each row is $128\times 5 = 640\,\mu$s plus overheads). This pattern is shown in, e.g., Fig.~2 of \citet{2020PASP..132g4504F}, and maps $1/f$ noise into a ``banding'' pattern (Fig.~\paperifignoiseonef\  of Paper I). Since our input $1/f$ noise fields are normalized to unit variance per logarithmic range in $f$, the appropriate normalization for the $1/f$ noise fields is
\begin{equation}
A\epsilon^2|_{1/f~\rm noise} = \frac{1}{\nu_{\rm SE}^2 \Omega_{\rm psf}/s_{\rm in}^2} \frac{f_{\rm knee}}{\Delta f_{\rm band}} \frac{\sigma^2_{\rm read}}{\sigma^2_{\rm tot}},
\end{equation}
where $\Delta f_{\rm band}$ is the bandwidth for the white noise (equal to half the sampling rate, so 100 kHz for {\slshape Roman}); and $\sigma^2_{\rm read}/\sigma^2_{\rm tot}$ is the fraction of the noise variance coming from read noise (as opposed to Poisson noise).

For both types of noise, we first attempted to mesure at single-epoch signal-to-noise ratio $\nu_{\rm SE}=10$; if the shape measurement did not converge (see Table~\ref{tab:noise_bias} for statistics on how often this happened) then we computed $A\epsilon^2$ using $\nu_{\rm SE}=10^{1.5}=31.6$.

\begin{figure*}
    \centering
    \includegraphics[width=\textwidth]{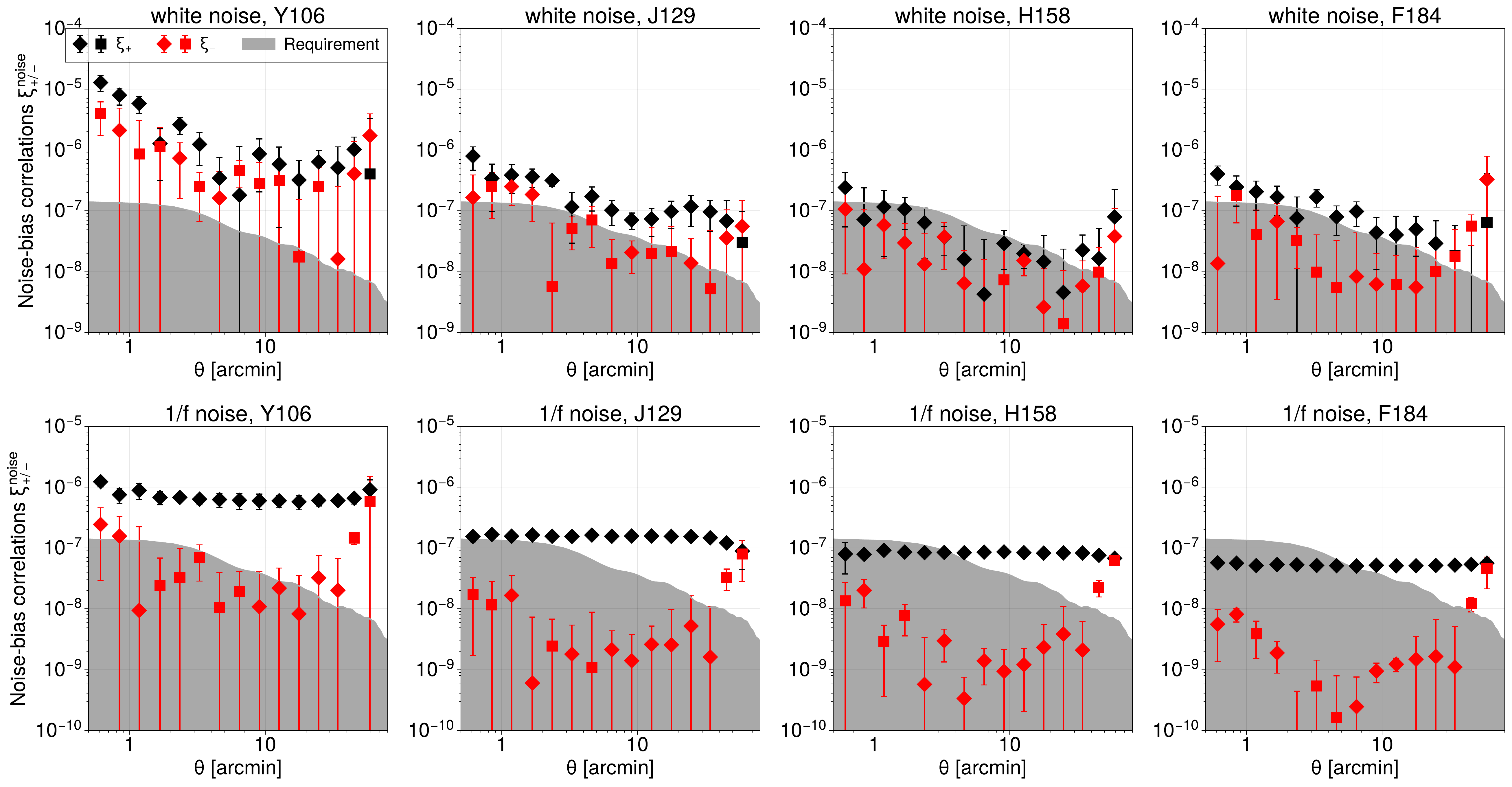}
    \caption{\label{fig:2pcf_noise_bias} Auto-correlations of the noise bias ($\Delta g$) estimated with white noise (\textbf{\emph{top}}) and 1/f (\textbf{\emph{bottom}}) noise models for all the four bandpasses are shown. ``diamond'' marker is the positive signal while ``square'' is the negative signal but taken absolute value. These are normalized to single-epoch signal-to-noise ratio $\nu_{\rm SE}=10$ and (for the $1/f$ noise case) a knee frequency of $f_{\rm knee}=1$\,kHz. Note that white noise biases generally scale as $\propto 1/\nu_{\rm SE}^2$ and $1/f$ noise biases scale as $\propto f_{\rm knee}/\nu_{\rm SE}^2$; this means that their correlation functions scale as $\propto 1/\nu_{\rm SE}^4$ and $\propto f_{\rm knee}^2/\nu_{\rm SE}^4$, respectively. These are the ``raw'' biases, with no attempt at mitigation. }
\end{figure*}

We report the noise-induced additive bias ($c_1$, $c_2$) for different noise models applied on the noise-less shapes of the {\sc GalSim}-made injected stars in Table \ref{tab:noise_bias}. As can be seen, the magnitude of additive biases surpasses the total additive systematic budget of $2.7 \times 10^{-4}$ according to the SRD. Furthermore, we correlated $\Delta g$ (which is the effect of noise in object ellipticity), and as can be seen in Fig.~\ref{fig:2pcf_noise_bias} noise-induced additive bias does show correlations over many scales. Hence, in the future, there will be a need to correct for these noise biases with a more comprehensive set of noise images if {\sc Imcom} is used for image processing for {\slshape Roman}.
Possible correction schemes range from numerical approaches based on shearing a noise field (as done in {\sc Metacalibration}; \citealt{2017ApJ...841...24S}); subtraction based on second derivatives of the weighted shape estimator \citep{2022arXiv220810522L}; or using analytic models such as Appendix~\ref{app:noise-additive} (presumably allowing the overall normalization to be empirically determined, as suggested in \citealt{2002AJ....123..583B}). It may be best in a cosmology analysis to implement more than one of these approaches as an internal check. It is clear that for the current set of {\sc Imcom} settings, the noise-induced biases are larger than residual PSF biases, suggesting that we might benefit from tuning the balance between leakage and noise in the future.

\begin{table*}
    \centering
    \caption{\label{tab:noise_bias} The mean values of the noise-induced additive bias ($c_1$ and $c_2$) for various input noise models and bandpasses. These are normalized to single-epoch signal-to-noise ratio $\nu_{\rm SE}=10$ and knee frequency $f_{\rm knee}=1$\,kHz, and are before any mitigations. (A very small fraction of objects was assigned $\nu_{\rm SE}=31.6$ (if {\tt galsim.hsm} did not converge for $\nu_{\rm SE}=10$) and re-normalized to $\nu_{\rm SE}=10$. These are shown in the table and are out of 54\,597 stars in total.) The errors calculated here are the standard error of the mean. The last two columns show predicted bias according to the formulae in Appendix~\ref{app:noise-additive} and the measured noise power spectra from Section~\ref{sec:noise_corr}.
    }
    \begin{tabular}[width=\columnwidth]{ccrrcrcrr}
\hline\hline
      Noise models & Bands & \multicolumn2c{Measured bias} & & \multicolumn1c{Number with} & & \multicolumn2c{Predicted bias} \\
      & & $c_1 \times 10^{4}$~~ & $c_2 \times 10^{4}$~~ & & \multicolumn1c{$\nu_{\rm SE}=31.6$} & & $c_1 \times 10^{4}$ & $c_2 \times 10^{4}$ \\ \hline
    white noise & Y106 & 5.44 $\pm$ 1.33 & $-$6.02 $\pm$ 1.25 & & 867~~~~~ & & 4.77 & $-5.01$ \\
    white noise & J129 & $-$0.81 $\pm$ 0.42 & $-$3.10 $\pm$ 0.40 & & 40~~~~~ & & $-0.81$ & $-1.89$ \\
    white noise & H158 & 1.24 $\pm$ 0.30 & $-$0.59 $\pm$ 0.30 & & 29~~~~~  & & 0.40 & $-0.51$ \\
    white noise & F184 & 1.14 $\pm$ 0.46 & 0.71 $\pm$ 0.46 & & 21~~~~~ &  & 0.26 & 0.80 \\ \hline
    $1/f$ noise & Y106 & 7.68 $\pm$ 0.38 & 1.42 $\pm$ 0.38 & & 44~~~~~ & & 5.62 & 0.62 \\
    $1/f$ noise & J129 & 3.80 $\pm$ 0.15 & 0.78 $\pm$ 0.08 & & 15~~~~~ &  & 2.34 & 0.54 \\
    $1/f$ noise & H158 & 2.87 $\pm$ 0.10 & 0.16 $\pm$ 0.06 & & 9~~~~~  & & 1.49 & 0.11 \\
    $1/f$ noise & F184 & $-$2.19 $\pm$ 0.08 & $-$0.56 $\pm$ 0.08 & & 9~~~~~ & & $-1.10$ & $-0.26$ \\
\hline\hline
    \end{tabular}
\end{table*}

We also show in Table~\ref{tab:noise_bias} the additive noise biases predicted by second-order biasing theory for Gaussian model objects Eq.~(\ref{eq:noisebias-g}), derived in Appendix~\ref{app:noise-additive}. (Note that all radii and power spectra must be scaled to the output pixel scale $s_{\rm out}$ in order to use that result.) The predicted biases are in the correct direction where significant and have approximately the right magnitude (the two largest biases in the table are greater than the prediction by 20\% and 37\%, although only the latter is statistically significant). The remaining discrepancy may be due to the assumption of a Gaussian profile used in the analytic bias prediction.

\section{Synthetic wide band images}
\label{sec:wide}

Synthetic wide band images, created by stacking several standard-width filters, are often used as an intermediate product in weak lensing analyses (e.g., $r+i+z$ in DES \citep{2018ApJS..239...18A, 2021ApJS..255...20A}). They can be used for making a deep detection image, or as a reference for forced photometry in individual filters for photometric redshifts. (This could include filters on other observatories, e.g., Vera Rubin Observatory data if one were to make a Y106+J129+H158+F184 image with {\slshape Roman}.) One could also make a shape catalog from the combined Y106+J129+H158+F184 imaging; such a catalog would have the advantages of providing a deeper catalog and reducing noise biases,\footnote{Noise biases usually scale as $\nu^{-2}$, where $\nu$ is the significance; so if one were to average $N$ exposures and then measure the ellipticity of a galaxy, the noise bias is a factor of $N$ smaller than if one measures the ellipticity in each exposure and takes the average. The calculation gets somewhat more complicated if one is considering different filters with different $\nu$, but the result that noise bias is reduced in the combined image should hold for most galaxy SEDs.} but the disadvantage that with only one catalog using all the data one can only measure a shape auto-correlation (we would not have the option of cross-correlating two different versions of the shape catalog from different data). It is also possible that one would use both the single-band and the synthetic wide band shape catalogs to test for different systematics --- for example, the individual bands allow for cross-correlations to test for PSF systematics associated with the tiling pattern (which is different in each filter; \citealt{2015arXiv150303757S}), but the wide-band image has lower noise bias.

We have built a set of wide band images in post-processing as follows. First, the output images in each filter $a$ are smoothed to a common output PSF $\Gamma_{\rm out}$ by convolving with a kernel $K_a$. Ideally, we want the coadd PSF $\Gamma_a$ convolved with this kernel to be the output PSF, $\Gamma_a\otimes K_a\approx \Gamma_{\rm out}$; in practice, we do this by writing
\begin{equation}
\tilde K_a({\bmath u}) = \frac{\tilde \Gamma^\ast_a({\bmath u})\tilde \Gamma_{\rm out}({\bmath u})}{|\tilde \Gamma_{\rm out}({\bmath u})|^2 + \epsilon^2}.
\end{equation}
Without the $\epsilon$ term, this kernel would exactly transform the PSF $\Gamma_a$ into $\Gamma_{\rm out}$; but we set $\epsilon = 10^{-4}$ in the denominator to avoid division by a near-zero quantity. The kernel is clipped to a $201\times 201$ output pixel region. We choose the output PSF $\Gamma_{\rm out}$ to be an Airy disc convolved with a Gaussian as in Paper~I; the Airy disc parameters are $\lambda/D = 0.112$ arcsec (the size for J129), obscuration 0.31, and the Gaussian has a scale length of $\sigma=0.1051$ arcsec (the largest of the Gaussian widths used in Paper~I). The full width at half maximum of this output PSF is 0.287 arcsec.

\begin{figure*}
\includegraphics[width=6in]{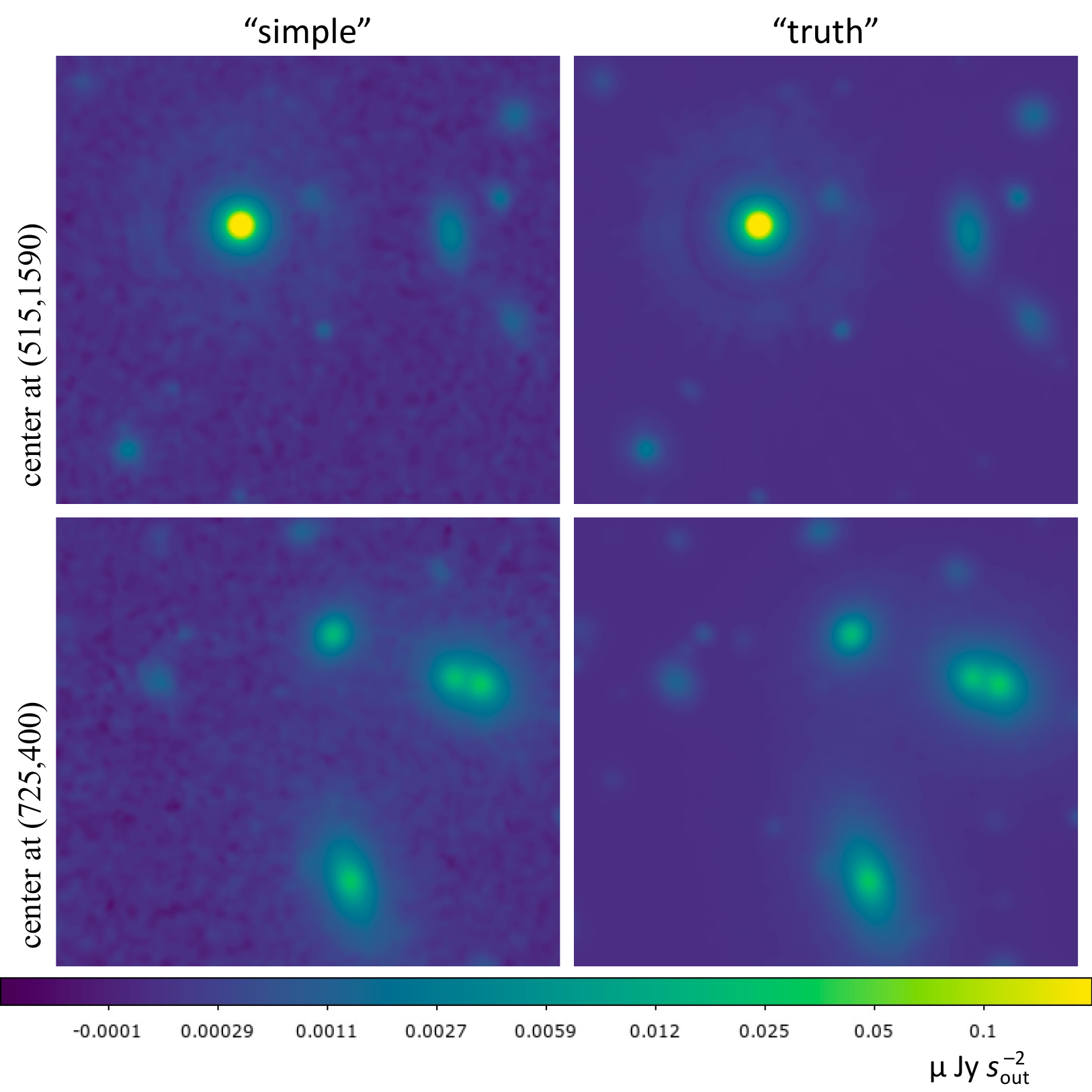}
\caption{\label{fig:wide}Two examples of fields in the synthetic wide band images (Sec.~\ref{sec:wide}), from block (0,14). The left column shows the ``simple'' detector model, the right column shows ``truth'' (the noiseless image). The top row contains a bright star that saturates on the color scale (reaching 0.90 $\mu$Jy $s_{\rm out}^{-2}$ at its center), with a log stretch to show the wings of the output PSF. Both images show a $450\times 400$ output pixel ($11.25\times 10$ arcsec) region.}
\end{figure*}

Next, the images are added with some weights:
\begin{equation}
H_{\rm out}(x,y) = \sum_a w_a C_a \sum_{\Delta x,\Delta y} K_a(\Delta x,\Delta y) H_a(x-\Delta x,y-\Delta y),
\end{equation}
where $H_a$ is the input image in band $a$, and $H_{\rm out}$ is the output image. The normalization factors $C_a$ convert the input units (electrons per $s_{\rm in}^2$ per exposure) into surface brightness units ($\mu$Jy $s_{\rm out}^{-2}$, or $\mu$Jy per output pixel) using the effective area curves provided by the {\slshape Roman} project.\footnote{https://roman.gsfc.nasa.gov/science/RRI/Roman\_effarea\_20210614.xlsx} The weights (summing to $\sum_aw_a=1$) are proportional to the inverse square depths given by \citet{2019arXiv190205569A}: 0.294, 0.323, 0.294, and 0.089 for the Y106, J129, H158, and F184 bands respectively. Examples of the synthetic wide images are shown in Fig.~\ref{fig:wide}.

\begin{figure*}
    \includegraphics[width=5.75in]{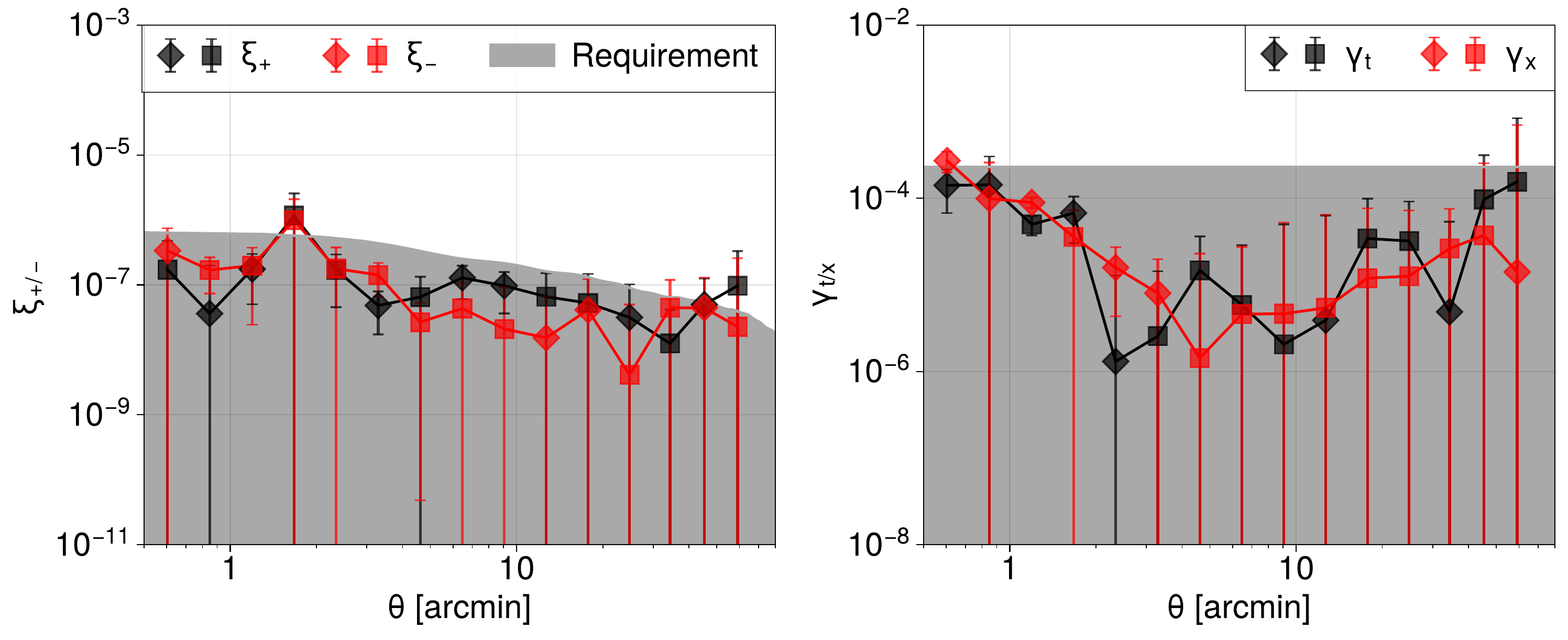}
    \caption{\label{fig:wide_corr}Shape-shape (\textbf{\emph{Left}}) and shape-position (\textbf{\emph{Right}}) correlation functions of stars observed in the synthetic wide band images. The grey region shows the estimated requirement on the signals from the PSF ellipticity additive error requirement in the SRD. ``diamond'' marker is the positive signal while ``square'' is the negative signal but taken absolute value.}
\end{figure*}

These images are then used again to extract stars to measure their properties, especially the two-point correlation functions. We measure the centroids and shapes of the stars using adaptive moments, and Figure~\ref{fig:wide_corr} shows their shape-shape and shape-position correlation functions. Although these signals suggest that the star ellipticity correlations are within the requirement defined in SRD, we do not see an improvement in removing the residual contamination compared to the cases with individual bandpasses shown in Fig.~\ref{fig:shear-shear}. 

\section{Summary and Discussion}
\label{sec:discussion}

In this paper, we have analyzed the coadded images that are produced in Paper~I. The layers that were simulated in Paper~I include the {\em simulated sky image} produced in \citet{2022arXiv220906829T}, the {\em injected point sources} ($\delta$-function convolved with {\slshape Roman} PSF), and two types of {\em noise fields} (white noise and 1/f noise). We have measured the following statistics on these images that are relevant to better characterize the weak lensing shear systematics:
\begin{itemize}
    \item the power spectra of coadded white noise and $1/f$ noise, as measured in the output images;
    \item one point statistics of magnitude, astrometric error, 2nd moments (shape and size), and 4th order moments of the injected stars (an idealized case with no blending or chromaticity effects) and stars in the coadded sky image; and 
    \item two-point statistics (shape-shape $\xi_{\pm}$ and position-shape $\gamma_{\rm t/x}$) of shapes and/or positions of the same sources.
\end{itemize}

If {\slshape Roman} images are to be processed using {\sc Imcom}, it is essential to understand the implications of noise biasing on measurements of galaxy shapes. Noise power spectra are directly involved in the calculation of shape measurement bias (Appendix~\ref{app:out-noise}), so we first investigated the power spectra of the coadded white and $1/f$ noise fields. In both cases, the noise power spectra of the coadded images decline to zero beyond $\sim 5$--6 cycles arcsec$^{-1}$, depending on the band. Specific features in the 2-dimensional and azimuthally averaged 1-dimensional power spectra indicate correlations in the noise fields that will impact shape measurements. We find that the most prominent features in the noise power spectra come from: band limits on the output PSFs, input image pixel positions, and the postage stamp boundaries and angular size. By comparing with analytic models of the expected 1-dimensional power spectra for well-sampled images (following derivations found in Appendix~\ref{app:out-noise}), we recover some of the expected qualitative features, but the noise power spectra are generally above the simple expectation. This highlights the importance of using full simulations of the image processing steps to predict the output noise and associated biases, rather than using simple formulae appropriate for well-sampled data. Each filter performs slightly differently, with the increase in power spectrum relative to the simplistic expectation being worst in Y106. This is unsurprising since it has the worst coverage and sampling of all the filters (see Paper~I for further discussion of coverage regions).

We have additionally presented the average astrometric offset from input catalog positions and the shape and size deviation from target PSF measured on injected stars (coadded PSF), and compared them to the PSF requirements documented in the {\slshape Roman} SRD. We have shown (Fig.~\ref{fig:starstatfig_v2}) that they are well within the requirements in a root-mean-square sense in the four bandpasses in the Reference survey design (Y106, J129, H158, F184). The {\sc Imcom} algorithm, although much more computationally expensive, has yielded large gains in output PSF quality relative to the current ``industry standard'' for combining space images ({\sc Drizzle}).

Since what matters to weak lensing studies in the end is the level of contamination each systematic effect has in shape correlations, we have measured two-point correlation functions of shapes (2nd moments: Fig.~\ref{fig:shear-shear}) of the injected and simulated (sky image) stars and 4th order moments (Fig.~\ref{fig:fourth_correlation}) of injected stars. While the PSF shape-shape correlations are well below the mission requirement for all the bandpasses, the 4th moment-4th moment correlations in F184 is on the verge of falling short (other 4th moment correlations are not of concern).

We finally investigated the additive bias imprinted on the shapes of the injected stars by both white and $1/f$ noise in the input images, by adding the appropriate noise fields in and re-measuring the shapes. We find biases that exceed {\slshape Roman} requirements when scaled to signal-to-noise ratio per observation $\nu_{\rm SE}=10$, indicating that noise bias will have to be corrected in {\slshape Roman} analyses (as is already planned for {\slshape Roman} and other weak lensing projects). Appendix~\ref{app:noise-additive} develops an analytic model for the noise bias, which we find is in agreement with the observed trends.

While this set of simulations represents a major step toward use of the {\sc Imcom} algorithm to the {\slshape Roman} weak lensing program, there are several more studies that should be carried out prior to implementation in an operational pipeline. These include:
\newcounter{SList}
\begin{list}{\arabic{SList}.\ }{\usecounter{SList}}
\item {\em Computational efficiency}: The existing implementation is computationally expensive, and would require $\sim 10^9$ CPU-hours to run on the full baseline HLIS survey if there were no speed-ups, whereas {\sc Drizzle} would require $\sim 10^5$ CPU hours (extrapolated from the implementation in \citealt{2022arXiv220906829T}). On the algorithm side, one could investigate optimizing postage stamp and padding parameters (see Paper~I), or re-arranging the linear algebra operations in {\sc Imcom} if we can search over a limited range in the Lagrange multiplier $\kappa$.\footnote{The variable $\kappa$ is the Lagrange multiplier that controls the relative weighting of leakage and noise in the optimization of the coaddition matrix ${\mathbfss T}$; see \citet{2011ApJ...741...46R}. In particular, with a limited range one could avoid the expensive eigendecomposition step in favor of matrix inversions.} On the hardware/firmware side, one could make use of ongoing advances in graphics processing unit-accelerated linear algebra.
\item {\em Extended source injection}: The tests with injected stars in this paper used point sources since these are a ``stress test'' for undersampling effects. However, we also want to implement grids of extended sources so that we can test {\sc Metacalibration} \citep{2017arXiv170202600H, 2017ApJ...841...24S} or analytic differentiation \citep{2022arXiv220810522L} techniques wrapped around an algorithm that operates on an {\sc Imcom} coadd.
\item {\em Error propagation}: We would like to study propagation of astrometric errors, relative flux calibration between images, and PSF model errors through the image coaddition pipeline. This will involve the insertion of specialized layers with these types of errors injected.
\item {\em Laboratory noise fields}: The analysis of the laboratory noise fields from this project, and an assessment of their implications for additive noise bias, is ongoing. Additional noise fields from the focal plane level and Wide Field Instrument level tests will be incorporated as they become available.
\item {\em Poisson noise bias corrections}: While {\sc Imcom} coaddition is a linear operation on the input pixels, shape measurement algorithms are non-linear operations on the data and hence are subject to noise biases. The current implementation allows one to coadd simulated noise relatizations, and therefore one can generate simulated output noise, as needed by tools such as Deep Metacalibration \citep{2022arXiv220607683Z}.\footnote{Note that Deep Metacalibration explicitly avoids the need to apply a shear to a ``wide survey'' noise field.} Such realizations would also allow a Monte Carlo evaluation of the noise bias terms in twice-differentiable shape measurement algorithms (e.g., \citealt{2022MNRAS.511.4850L, 2022arXiv220810522L}). The performance of the method has not been tested on simulations with source Poisson noise. But further work will be needed to address biases arising from self-Poisson noise of the source galaxies \citep[e.g., Appendix D of][]{2022arXiv220810522L}.
\item {\em Chromatic effects}: When running image combination in mosaic mode, one has to choose a ``reference'' intraband SED in order to have a well-defined PSF, but of course the sources in the field have a range of SEDs. Chromatic terms in the PSF will arise from diffraction, dispersion in the filter substrate, chromatic wavefront from mirror, filter, and detector coatings, and depth-dependent absorption effects in the detectors \citep{2020JATIS...6d6001M}. Additionally, there is a field-dependent filter bandpass due to angle of incidence effects, which results in objects redder than the reference SED having a normalization that is larger for input exposures near the field center and smaller for input exposures near the corners of the field. We plan to test how these chromatic effects propagate through {\sc Imcom} and assess how they should be corrected.
\item {\em Deep fields}: Like other weak lensing surveys, the {\slshape Roman} HLIS requires deep fields for photo-$z$, selection, and noise bias calibration. Some shear calibration algorithms also require deep fields \citep{2022arXiv220607683Z}. The {\sc Imcom} algorithm will run on deep fields, but since the current version has ${\mathcal O}(n^3)$ (where $n$ is the number of input pixels), a deep survey with $10$ as many exposures but only 1\% of the area of the wide survey actually requires $10^3\times 0.01 = 10$ times as many operations to run {\sc Imcom} as the wide survey if done by brute force. One solution to mitigate this is to coadd subsets of the deep field exposures to obtain full sampling at a common PSF, and then do a pixel-by-pixel coadd, but other options should be investigated for feasibility and performance. Another opportunity for the deep fields, since the observations include large dithers, would be to mitigate field-dependent bandpass effects. This would involve extending the linear algebra formalism in \citet{2011ApJ...741...46R} to simultaneously reconstruct both an image through the ``mean'' bandpass in each filter and a ``first principal component'' (PC1) bandpass.\footnote{Mathematically, this would involve turning the optical transfer function $\tilde G_i({\bmath u})$ into a length 2 vector for the sensitivity to the mean and PC1 bandpasses. The ${\mathbfss A}$ and ${\mathbfss B}$ matrices would involve dot products of these vectors as well as integrals over the Fourier plane.}
\item {\em Other survey strategies}: The Reference survey strategy was developed to support hardware trades for the {\slshape Roman} mission, and the actual survey design may be different (see, e.g., \citealt{2021MNRAS.507.1514E} for one proposal). Simulations similar to those in this series of papers should be carried out for the possible alternative survey designs to ensure the data will be usable.
\end{list}

\section*{Acknowledgements}

During the preparation of this work, M.Y. and M.T. were supported by NASA contract 15-WFIRST15-0008 as part of the Roman Cosmology with the High-Latitude Survey Science Investigation Team and by NASA under JPL Contract Task 70-711320, ``Maximizing Science Exploitation of Simulated Cosmological Survey Data Across Surveys.'' E.M. and C.H. were also supported by NASA contract 15-WFIRST15-0008, Simons Foundation  grant 256298, and David \& Lucile Packard Foundation grant 2021-72096. Work at Argonne National Laboratory was supported under the U.S. Department of Energy contract DE-AC02-06CH11357. This research used resources of the Argonne Leadership Computing Facility, which is supported by DOE/SC under
contract DE-AC02-06CH11357. R.M. was supported by a grant from the Simons Foundation (Simons Investigator in Astrophysics, Award ID 620789).

We thank Kirsten Casey for assistance with the azimuthally averaged power spectra, and Eric Huff and Chaz Shapiro for comments on the derivation in Appendix~\ref{app:noise-additive}. We thank Paul Martini, Chun-Hao To, Peter Taylor, and David Weinberg for feedback on the methodology and analysis choices for this project.

The detector data files used in the {\slshape Roman} image simulations are based on data acquired in the Detector Characterization Laboratory (DCL) at the NASA Goddard Space Flight Center. We thank the personnel at the DCL for making the data available for this project. Computations for this project used the Pitzer cluster at the Ohio Supercomputer Center \citep{Pitzer2018} and the Duke Compute Cluster.

This project made use of the numpy \citep{2020Natur.585..357H} and astropy  \citep{2013A&A...558A..33A,2018AJ....156..123A, 2022ApJ...935..167A} packages.
Some of the figures were made using ds9 \citep{2003ASPC..295..489J} and {\sc Matplotlib} \citep{Hunter:2007}.
Some of the results in this paper have been derived using the healpy and HEALPix package \citep{2005ApJ...622..759G, 2019JOSS....4.1298Z}.

\section*{Data Availability}

The codes for this project, along with sample configuration files and setup instructions, are available in the two GitHub repositories:
\begin{itemize}
\item https://github.com/hirata10/furry-parakeet.git (postage stamp coaddition)
\item https://github.com/hirata10/fluffy-garbanzo.git (mosaic driver)
\end{itemize}
The version of the code used for this project is in tag 0.1.

\bibliographystyle{mnras}
\bibliography{mainbib}

\appendix

\section{Models for the output noise power spectra in the absence of sampling effects}
\label{app:out-noise}

This appendix derives the analytic predictions for output noise power spectra in the absence of sampling effects, as used in Section \ref{sec:noise_corr}. There are two cases for the input noise: white noise, and $1/f$ noise.

\subsection{Input white noise}

To calculate the expected output power spectrum for white noise, we start by writing the input and output images (\{$I_{a}\}_{a=0}^{N_{\rm in}-1}$ and $H_{\rm out}$) as convolutions of the input and output PSFs $G_{\rm in}$ and $\Gamma_{\rm out}$ with the sky scene $f({\bmath r})$:
\begin{equation}
I_{a}(\vec{r}) = [G_{\rm in} \ast f]({\bmath r}) 
{\rm ~~and~~}
H_{\rm out}({\bmath r}) = [\Gamma_{\rm out} \ast f]({\bmath r}).
\end{equation}
where the input PSF $G_{\rm in}$ is the Airy disc convolved with a top hat function, $a$ is the index of the input PSF, and the target PSF $\Gamma_{\rm out}$ is the Airy disc convolved with a Gaussian. Taking the Fourier transform and combining these equations for each input image $j$ gives:
\begin{equation}
    \tilde{H}_{{\rm out}} = \frac{\tilde{\Gamma}_{\rm out}({\boldsymbol\umag})}{\tilde{G}_{\rm in}({\boldsymbol\umag})} \sum_{j=0}^{N_{\rm in}-1} \frac{\tilde{I}_{a,j}({\boldsymbol\umag})}{N_{\rm in}},
\end{equation}
where ${\boldsymbol\umag}=(u,v)$ is the wave vector. For independent input images, the power spectrum of the output is
\begin{equation}
    {P}_{\rm out}(\vec{k}) = \Bigg|\frac{\tilde{\Gamma}_{\rm out}({\boldsymbol\umag})}{\tilde{G}_{\rm in}({\boldsymbol\umag})}\Bigg|^2 \frac{{P}_{\rm in}({\boldsymbol\umag})}{N_{\rm in}}.
\end{equation}
Here the Airy disc components of the PSFs cancel each other out when we take the ratio $\tilde{\Gamma}_{\rm out}/\tilde{G}_{\rm in}$, so this ratio is simply the ratio of the Gaussian to the sinc (Fourier transform of the top-hat). Following {numpy} convention and Paper~I, we use the definition of the sinc function ${\rm sinc}(\xi)=\sin{\pi\xi}/\pi\xi=j_0(\pi\xi)$, where $j_0$ is the zeroth order spherical Bessel function.
Then
\begin{equation}
    {P}_{\rm out}({\boldsymbol\umag}) = \Bigg| \frac{{\rm e}^{-2\pi^2\sigma^2\umag^2}}{\text{sinc}(s_{\rm in}u) \,\text{sinc}(s_{\rm in}v)} \Bigg|^2
    \frac{P_{\rm in}({\boldsymbol\umag})}{N_{\rm in}}.
\end{equation}
In this case, we are interested in input white noise with unit variance in pixels of area $s_{\rm in}^2$, so the input power spectrum is $P_{\rm in}({\boldsymbol\umag})=s_{\rm in}^2$.

We now taking an angular average over the direction of the Fourier mode $\varphi \in [0, 2\pi)$ at fixed magnitude $\umag$. In this case, $u=\umag\cos\varphi$ and $v=\umag\sin\varphi$. After a lengthy but straightforward calculation, we get
\begin{equation}
    \Big\langle {P}_{\rm out}({\boldsymbol\umag})\Big\rangle_\varphi = \frac{s_{\rm in}^2}{N_{\rm in}} \text{exp} \left[ 4 \pi^2 \left(\frac{s_{\rm in}^2}{12} - \sigma^2\right)\umag^2 -\frac{\pi^4}{120} s_{\rm in}^4 \umag^4 + ... \right],
\label{eq:P-out-whitenoise}
\end{equation}
where the ``...'' denote higher terms in the Taylor expansion of $\umag$.
Note that the power spectrum has units of area since it is of an input field that is dimensionless by construction.
This is the analytic expectation for the white noise power spectrum used in the analysis in Section~\ref{sec:noise_corr} in this work.

\subsection{Input \texorpdfstring{$1/f$}{1/f} noise}

We now consider the case of input $1/f$ noise. To avoid clutter, we work initially in units of the input pixel scale (up through Eq.~\ref{eq:pmeas2}), and then transform units at the end.

The input $1/f$ noise fields are generated as 1D time-ordered data (TOD), and re-formatted into a 2D image.
Thus for the prediction of an output noise power spectrum for input $1/f$ noise, we start with the power spectrum as a function of frequency on the 1D pixel axis,
\begin{equation}
P_{\rm TOD}(f) = 
\begin{cases}
\frac{1}{2|f|} & \text{if } \frac{1}{N^2}<|f|<\frac{1}{2} \\
0 & \text{otherwise}
\end{cases},
\label{eq:1dpf}
\end{equation}
where $\frac12$ is the Nyquist frequency (in units where the time to sample each pixel is unity) and $1/N^2$ is the cutoff for a sequence of $N^2$ pixels.
Any two pixels separated by some time $t$ have a correlation function
\begin{equation}
    \xi_{\rm TOD}(t) = \int P_{\rm TOD}(f)e^{2\pi\,{\rm i}ft}\,{\rm d}f 
    = \int_{f_{min}}^{1/2}\frac{1}{f}\cos(2\pi\,ft) \,{\rm d}f,
\end{equation}
where we have substituted $P_{\rm TOD}(f)$, used the fact that the $f<0$ range contributes the conjugate of the $f>0$ range, and set $f_{max}=1/2$ (the Nyquist frequency).

Pixels separated by $t$ in the time series are separated by some $(\Delta x, \Delta y)$ in the input data. The 1D (TOD) and 2D correlation functions are thus of the form $\xi_{\rm 2D}(\Delta x, \Delta y)$ and $\xi_{\rm TOD}(\Delta x + N\Delta y)$. Summing over $(\Delta x, \Delta y)$ gives us a 2D power spectrum (in Fourier space):
\begin{eqnarray}
{P}_{\rm 2D}(u,v) &=& \sum_{\Delta x,\Delta y}\xi_{\rm 2D}(\Delta\,x,\Delta\,y)e^{2\pi\,{\rm i}(u\Delta x+v\Delta y)} 
\nonumber \\
 &=& \sum_{\Delta x,\Delta y} \xi_{\rm TOD}(\Delta x+N\Delta y) e^{2\pi\,{\rm i}(u\Delta x+v\Delta y)}
 \nonumber \\
 &=& \sum_{\Delta y} e^{2\pi{\rm i} v\Delta y} \Big[ \sum_{\Delta x}\xi_{\rm TOD}(\Delta x+N\Delta y) 
 \nonumber \\ && ~~~~\times 
 e^{2\pi\,{\rm i}u(\Delta x+N\Delta y)}\Big]e^{-2\pi\,{\rm i}uN\Delta y}.
\end{eqnarray}
In the last line we have factored out an exponential so that we may substitute, in analogy with the first line, 
\begin{equation}
{P}_{\rm TOD}(u)=\sum_{\Delta x}\xi_{\rm TOD}(\Delta x+N\Delta y)e^{2\pi{\rm i}u(\Delta x+N\Delta y)}.
\end{equation}
This gives the 2D power spectrum of the image as a function of the 1D power spectrum of the time-ordered data:
\begin{equation}
    {P}_{\rm 2D}(u,v) = \sum_{\Delta y} e^{2\pi{\rm i}v\Delta y} P_{\rm TOD}(u) e^{-2\pi{\rm i} uN\Delta y},
\end{equation}
which, using the exponential form definition of the $\Sha$-function $\Sha(x)=\sum_{n=-\infty}^{\infty}e^{2\pi{\rm i}nx}$ (the ``Shah'' or Dirac Comb function), becomes:
\begin{equation}
    {P}_{\rm 2D}(u,v) = \Sha(v-uN) \tilde{P}_{\rm TOD}(u).
\label{eq:P2D-Sha}
\end{equation}

Equation~(\ref{eq:P2D-Sha}) is general in the case that one considers the full plane, extending to infinity. However, the presence of the $\Sha$-function (and hence $\delta$-function) presents a challenge to numerical computation. The solution is to ``smear out'' the $\delta$-function according to the uncertainty principle given that our detector array is finite in size. In a finite region, the measured power spectrum ${P}_{\rm meas}({\boldsymbol\umag})$ is related to the true power spectrum ${P}({\boldsymbol\umag})$ by
\begin{equation}
    {P}_{\rm meas}({\boldsymbol\umag}) = \frac{1}{A}\int\,{\rm d}^2{\boldsymbol\umag}'{P}({\boldsymbol\umag})|\tilde{W}({\boldsymbol\umag}-{\boldsymbol\umag}')|^2,
\end{equation}
where $A$ is the area of the space and $\tilde{W}$ is the window function (Fourier transform of the region measured; see discussion in Section 13.4 of \citealt{1992nrca.book.....P}).

We consider a rectangle of width $N$ (in the $x$-direction) and $N_y$ (in the $y$-direction) gives $A=NN_y$ and a window function
\begin{eqnarray}
|\tilde{W}(\Delta u,\Delta v)|^2 
\!\!\!\! &=& \!\!\!\! (NN_y)^2 \,{\rm sinc}^2(N\Delta u)\,{\rm sinc}^2(N_y\Delta v)
\nonumber \\
&\approx & \!\!\!\!
{\rm sinc}^2(N\Delta u)\, N^2 N_y\delta(\Delta\,v).
\end{eqnarray}
Applying this to our 2D power spectrum for $1/f$ noise gives
\begin{eqnarray}
    {P}_{\rm 2D,meas}(u,v)&=& \frac{1}{NN_y}\int\,{\rm d}\Delta u\,{\rm d}\Delta v\, \Sha(v'-u'N){P}_{\rm TOD}(u')
    \nonumber \\
    && ~~~~ \times {\rm sinc}^2(N\Delta u)\, N^2N_y\delta(\Delta\,v)\text{ },
\end{eqnarray}
where we have defined $u'=u-\Delta u$ and $v'=v-\Delta v$. Performing the integral over $\Delta v$ thus takes $v'\rightarrow v$, resulting in
\begin{eqnarray}
    {P}_{\rm 2D,meas}(u,v) \!\!\!\! &=& \!\!\!\! N \int {\rm d}\Delta u\,  \Sha(v-uN+N\Delta u) 
    \nonumber \\ && ~~~~
    \times {P}_{\rm TOD}(u-\Delta u) \,{\rm sinc}^2(N\Delta u).~~~~
\end{eqnarray}
The $\Sha$-function can equivalently be written as $\Sha(x)=\sum_{n=-\infty}^{\infty}\delta(x-n)$, which will allow us to trivially perform the integral over $\Delta u$ as well:
\begin{equation}
{P}_{\rm 2D,meas}(u,v) 
= \sum_{j=-\infty}^{\infty} {P}_{\rm TOD}\left(\frac{v-j}{N}\right)\,{\rm sinc}^2(Nu+j-v).
\end{equation}
Finally, we substitute in our time-ordered power spectrum given for $1/f$ noise, Eq.~(\ref{eq:1dpf}), and find that the measured power spectrum for $1/f$ noise should be
\begin{equation}
    {P}_{\rm 2D,meas}(u,v) = \frac{N}{|v-j|} \,{\rm sinc}^2(Nu+j-v).
\end{equation}

To include this result in our 1D representations in Figure~\ref{fig:1dspectra}, we convert to polar coordinates $\umag = \sqrt{u^2+v^2}$ and $\varphi =\, {\rm arctan2}(v,u)$, and take the angular average:
\begin{eqnarray}
\left\langle{P}_{\rm 2D,meas}(\umag,\varphi)\right\rangle_{\varphi}
\!\!\!\! &=& \!\!\!\! \int_{0}^{2\pi} \frac{{\rm d}\varphi}{2\pi} \sum_{j=-\infty}^{\infty} {P}_{\rm TOD}\left(\frac{\umag\sin\varphi-j}{N}\right)
\nonumber \\ &&
\times {\rm  sinc}^2(N\umag\cos\varphi+j-\umag\sin\varphi).
\end{eqnarray}

We can restrict the range of $j$ since the TOD power spectrum has a Nyquist frequency, so we include only
\begin{equation}
\frac{-1}{2} < \frac{r\sin\varphi-j}{N} < \frac{1}{2}.
\end{equation}
The integral can be turned into a sum using $\varphi=2\pi\,K_\varphi/N_\varphi$, resulting in the final form we use to compute the expectations for the 1D $1/f$ noise power spectra:
\begin{eqnarray}
{P}_{\rm meas,2D\,ang.\,ave.}(\umag) \!\!\!\! &=& \!\!\!\! \frac{1}{N_\varphi} \sum_{K_\varphi=0}^{N_\varphi-1} \sum_{j=j_{\rm min}}^{j_{\rm max}} {P}_{\rm TOD}\left(\frac{\umag\sin\varphi-j}{N}\right)
\nonumber \\ &&
\times \,{\rm sinc}^2(N\umag\cos\varphi+j-\umag\sin\varphi).
\label{eq:pmeas2}
\end{eqnarray}
We note that this is in a single input image, in units where $s_{\rm in}=1$. To compare to the power spectra in Sec.~\ref{ss:noise-1D}, we must convert back to general units, include the factor of $1/N_{\rm in}$ for the input exposures, and include the square of ratios of modulation transfer functions to convert to the output images (as in Eq.~\ref{eq:P-out-whitenoise}). The result is
\begin{eqnarray}
    \Big\langle {P}_{\rm out}({\boldsymbol\umag})\Big\rangle_\varphi \!\!\!\! &=& \!\!\!\!
    \frac{1}{N_\varphi} \sum_{K_\varphi=0}^{N_\varphi-1} \sum_{j=j_{\rm min}}^{j_{\rm max}} {P}_{\rm TOD}\left(\frac{s_{\rm in}\umag\sin\varphi-j}{N}\right)
\nonumber \\ &&
\times \,{\rm sinc}^2(Ns_{\rm in}\umag\cos\varphi+j-s_{\rm in}\umag\sin\varphi)
    \nonumber \\ && \times
    \frac{s_{\rm in}^2}{N_{\rm in}} \text{exp} \left[ 4 \pi^2 \left(\frac{s_{\rm in}^2}{12} - \sigma^2\right)\umag^2 -\frac{\pi^4}{120}  s_{\rm in}^4 \umag^4 + ... \right].
\nonumber \\ &&
\label{eq:pmeas2-all}
\end{eqnarray}
We use $N=128$ as the width of the output channels \citep{2020JATIS...6d6001M}.

\section{Additive bias from anisotropic noise}
\label{app:noise-additive}

While we have used injected stars with and without noise added to estimate the additive noise bias in the main text of the paper (Section~\ref{subsec:noise_bias}), it is useful to have an analytic model in order to understand the orders of magnitude in the problem and the scaling behaviors. These analytic models, once calibrated with simulations, are also useful for setting requirements on knowledge of correlated noise --- in particular, they were used to determine how many dark images needed to be taken at each calibration epoch to measure read noise correlations in the Roman Design Reference Mission\footnote{RST-SYS-DESC-0073; \url{https://asd.gsfc.nasa.gov/romancaa/docs2/RST-SYS-DESC-0073-_D10.xlsx}}. The main objective of this appendix is to predict the additive bias $c$ for ellipticity of stars or of galaxies in terms of the noise power spectrum $P_{\rm N}(u,v)$. For simplicity, this appendix does so for the adaptive moments method \citep{2002AJ....123..583B}. A related previous study on noise biases in fitting elliptical Gaussians can be found in \citet{1997PASP..109..166C}.

We follow the description of biases in Appendix C of \citet{2004MNRAS.353..529H} and Appendix A of \citet{2012MNRAS.425.1951R}. This description applies to the biases in a least-squares fitting algorithm where the galaxy image $I({\bmath x})$ is fit by a model $J({\bmath x}|{\bmath p})$, where ${\bmath p}$ is a point in a parameter space with parameters $\{p^\alpha\}$ (with $N$ parameters). The metric for the least-squares fit is assumed to be an inner product, denoted by $\langle,\rangle$: that is, the ``energy functional'' that is minimized is
\begin{equation}
E = \langle I-J({\bmath p}), I-J({\bmath p})\rangle = \frac12 \int |I({\bmath x})-J({\bmath x}|{\bmath p})|^2\,d^2{\bmath x}.
\label{eq:E0}
\end{equation}
This definition guarantees the functional derivative
\begin{equation}
\frac{\delta}{\delta g({\bmath x})}\langle f,g\rangle = f({\bmath x}).
\end{equation}
We assume we are working with a well-sampled image (in the context of this paper, that means the output image generated by {\sc Imcom}) so that summations over pixels instead of integrals, and hence partial instead of functional derivatives, may also be used. The case considered here is more difficult than that in \citet{2004MNRAS.353..529H}, because in the \citet{2004MNRAS.353..529H} calculation the least squares fit is inverse-noise-weighted with respect to the true noise covariance matrix. In contrast, here we are interested in the case where the true noise covariance is {\em not} the weight used in the inner product of Eq.~(\ref{eq:E0}). Indeed, the true noise covariance may not be known exactly, and one of our purposes is to understand what impact errors in the noise model have on output shape measurements.

We further use subscripts on $J$ to indicate partial derivatives with respect to $p^\alpha$, and drop the arguments unless explicitly required for clarity: thus $J_\alpha = \partial J({\bmath x}|{\bmath p})/\partial p^\alpha$.

We first consider the general calculation of mean shifts and biases, and then proceed to the case of galaxy ellipticity.

\subsection{General results}

We note that the minimization of $E$ with respect to a parameter $p^\alpha$ leads to the minimization equation:
\begin{equation}
0 = \frac12\frac{\partial E}{\partial p^\alpha} = \langle I-J, J_\alpha\rangle.
\label{eq:min}
\end{equation}
This is a set of $N$ nonlinear equations for the $N$ parameters $\{p^\alpha\}_{\alpha=0}^{N-1}$ in terms of the image $I$. Clearly if $I = J({\bmath P})$ for some point ${\bmath P}$ in parameter space, then the solution to Eq.~(\ref{eq:min}) is that $p^\alpha=P^\alpha$.

This solution can be Taylor-expanded. We must first take the functional derivative of Eq.~(\ref{eq:min}) with respect to the image:
\begin{equation}
0 = J_\alpha({\bmath r}) + \left[ -\langle J_\beta,J_\alpha\rangle + \langle I-J,J_{\alpha\beta}\rangle \right] \frac{\delta p^\beta}{\delta I({\bmath r})},
\label{eq:D1}
\end{equation}
and then one more derivative:
\begin{eqnarray}
0 &=& J_{\alpha\beta}({\bmath r}) \frac{\delta p^\beta}{\delta I({\bmath s})} +
\Bigl[ -\langle J_{\beta\gamma},J_\alpha\rangle - \langle J_\beta,J_{\alpha\gamma}\rangle
- \langle J_\gamma,J_{\alpha\beta}\rangle 
\nonumber \\ &&
+ \langle I-J,J_{\alpha\beta\gamma}\rangle 
\Bigr] \frac{\delta p^\beta}{\delta I({\bmath r})} \frac{\delta p^\gamma}{\delta I({\bmath s})}
+ J_{\alpha\beta}({\bmath s}) \frac{\delta p^\beta}{\delta I({\bmath r})}
\nonumber \\ &&
+ \left[ -\langle J_\beta,J_\alpha\rangle + \langle I-J,J_{\alpha\beta}\rangle \right] \frac{\delta^2p^\beta}{\delta I({\bmath r})\delta I({\bmath s})}.
\label{eq:D2}
\end{eqnarray}
If we focus next on expanding around the point where $I=J$, then --- defining ${ H}^{\alpha\beta}$ to be the $N\times N$ matrix inverse of the symmetric matrix $\langle J_\beta,J_\alpha\rangle$ --- we find
\begin{equation}
\frac{\delta p^\alpha}{\delta I({\bmath r})} = H^{\alpha\beta} J_\beta({\bmath r})
\label{eq:D1e}
\end{equation}
and (after some simplification)
\begin{eqnarray}
\frac{\delta^2p^\beta}{\delta I({\bmath r})\delta I({\bmath s})} &=&
H^{\alpha\beta} H^{\gamma\delta} \frac{\partial}{\partial p^\alpha} \left[ J_{\delta}({\bmath r}) J_\gamma({\bmath s}) \right]
\nonumber \\ &&
- \left[ \langle J_{\delta\gamma},J_\alpha\rangle + \langle J_\delta,J_{\alpha\gamma}\rangle
+ \langle J_\gamma,J_{\alpha\delta}\rangle \right] 
\nonumber \\ && \times
H^{\alpha\beta} H^{\gamma\epsilon} H^{\delta\zeta}
J_\epsilon({\bmath r}) J_\zeta({\bmath s}).
\label{eq:D2e}
\end{eqnarray}

Now we suppose that there is some noise covariance matrix ${\mathbfss N}({\bmath r},{\bmath s}) = \langle \Delta I({\bmath r}) \Delta I({\bmath s})\rangle$. Then let us define the symmetric matrix
\begin{eqnarray}
Q_{\epsilon\zeta} \!\!\!\! &=& \!\!\!\! \int J_\epsilon({\bmath r}) J_\zeta({\bmath s}) {\mathbfss N}({\bmath r},{\bmath s})\,{\rm d}^2{\bmath r}\,{\rm d}^2{\bmath s}
\nonumber \\
&=&  \!\!\!\! \int \tilde J^\ast_\epsilon({\boldsymbol\umag}) \tilde J_\zeta({\boldsymbol\umag}) P_N({\boldsymbol\umag})\,d^2{\boldsymbol\umag},
\label{eq:NoiseQ.2}
\end{eqnarray}
where $P_{\rm N}({\boldsymbol\umag})$ is the noise power spectrum at wave vector ${\boldsymbol\umag}=(u,v)$.

The covariance of the model parameters, from Eq.~(\ref{eq:D1e}), is then
\begin{equation}
{\rm Cov}(p^\alpha,p^\beta) = H^{\alpha\gamma} H^{\beta\delta} Q_{\gamma\delta},
\label{eq:COV}
\end{equation}
and the bias is
\begin{equation}
\Delta p^\beta = \frac12\int {\mathbfss N}({\bmath r},{\bmath s}) \frac{\delta^2 p^\beta}{\delta I({\bmath r})\delta I({\bmath s})}\,{\rm d}^2{\bmath r}\,{\rm d}^2{\bmath s}.
\end{equation}
After simplifying using the idenitity (provable using the product rule)
\begin{equation}
\langle J_\delta,J_{\alpha\gamma}\rangle
+ \langle J_\gamma,J_{\alpha\delta}\rangle = \frac{\partial [{\mathbfss H}^{-1}]_{\gamma\delta}}{\partial p^\alpha},
\label{eq:prodrule}
\end{equation}
this becomes
\begin{equation}
\Delta p^\beta = \frac12 H^{\alpha\beta} \left[ \frac{\partial }{\partial p^\alpha} (H^{\gamma\delta} Q_{\gamma\delta})
-  \langle J_{\delta\gamma},J_\alpha\rangle H^{\gamma\epsilon} H^{\delta\zeta} Q_{\epsilon\zeta} \right],
\label{eq:BIAS}
\end{equation}
where from Eq.~(\ref{eq:prodrule}):
\begin{equation}
\langle J_{\delta\gamma},J_\alpha\rangle = \frac12 \left( \frac{\partial[{\mathbfss H}^{-1}]_{\alpha\delta}}{\partial p^\gamma}
+ \frac{\partial[{\mathbfss H}^{-1}]_{\alpha\gamma}}{\partial p^\delta}
- \frac{\partial[{\mathbfss H}^{-1}]_{\gamma\delta}}{\partial p^\alpha} \right).
\label{eq:JJ}
\end{equation}
Note the appearance of the bias tensor, as discussed in \citet{CoxSnell1968}; this is very much like a Christoffel symbol if ${\mathbfss H}$ is interpreted as a contravariant metric on parameter space. It appears in other contexts where we consider noise biases in photometry \citep[e.g.][]{2020AJ....159..165P}. The first term, involving the derivative of $H^{\gamma\delta}Q_{\gamma\delta}$, does not appear in the usual formula for the bias of a maximum likelihood estimator; it results essentially from the fact that the weight in the inner product $\langle,\rangle$ used to define the energy functional $E$ is {\em not} an inverse-noise weighting with the true noise. In particular, one can show that for white noise with $P_{\rm N}({\boldsymbol\umag})=A$ that ${\mathbfss Q} = A{\mathbfss H}^{-1}$, and hence $\partial(H^{\gamma\delta}Q_{\gamma\delta})/\partial p^\alpha=0$. But it needs to be taken into account in the present context, where we allow an arbitrary noise power spectrum but then fit objects with uniform weighting.

Equation~(\ref{eq:BIAS}) is additive in the noise correlation function, as is always the case for second-order biases.

\subsection{Application to ellipticities}

We now consider the case where the particular model of interest is the fitting of an elliptical Gaussian:
\begin{equation}
J({\bmath x}|{\bmath p}) = \frac{F}{2\pi\sqrt{\det\mathbfss M}} \exp\frac{-({\bmath x}-{\bmath x}_{\rm c})\cdot {\mathbfss M}^{-1}({\bmath x}-{\bmath x}_{\rm c})}{2},
\end{equation}
where the parameters are the flux $F$, the centroid $(x_c,y_c)$, and the three moments of the $2\times 2$ symmetric matrix ${\mathbfss M}$, usually written in the form
\begin{equation}
{\mathbfss M} = R^2 \left( \begin{array}{cc} 1+e_1 & e_2 \\ e_2 & 1-e_1 \end{array} \right).
\end{equation}
This is the \citet{2002AJ....123..583B} convention for ellipticities; the scale parameter $R$ was chosen to make ${\mathbfss M}$ simple, and is defined in terms of the principal axes by $R^2 = (a^{2} + b^{2})/2$. The Fourier transforms are
\begin{equation}
\tilde J({\boldsymbol\umag}|{\bmath p}) = Fe^{-2\pi i{\bmath x}_{\rm c}\cdot{\boldsymbol\umag}}e^{-2\pi^2{\boldsymbol\umag}\cdot{\mathbfss M}{\boldsymbol\umag}}
\label{eq:J-adaptive}
\end{equation}
(it is easiest to compute the inner products in Fourier space).

The derivatives of $J$ are
\begin{eqnarray}
\tilde J_F &=& {\rm e}^{-2\pi {\rm i}{\boldsymbol x}_{\rm c}\cdot{\boldsymbol\umag}}{\rm e}^{-2\pi^2{\boldsymbol\umag}\cdot{\bf M}{\boldsymbol\umag}},
\nonumber \\
\tilde J_{x_c} &=& -2\pi {\rm i}uF{\rm e}^{-2\pi {\rm i}{\boldsymbol x}_{\rm c}\cdot{\boldsymbol\umag}}{\rm e}^{-2\pi^2{\boldsymbol\umag}\cdot{\bf M}{\boldsymbol\umag}},
\nonumber \\
\tilde J_{y_c} &=& -2\pi {\rm i}vF{\rm e}^{-2\pi {\rm i}{\boldsymbol x}_{\rm c}\cdot{\boldsymbol\umag}}{\rm e}^{-2\pi^2{\boldsymbol\umag}\cdot{\bf M}{\boldsymbol\umag}},
\nonumber \\
\tilde J_R &=& -\frac{4\pi^2}R{\boldsymbol\umag}\cdot{\bf M}{\boldsymbol\umag} F{\rm e}^{-2\pi {\rm i}{\boldsymbol x}_{\rm c}\cdot{\boldsymbol\umag}}{\rm e}^{-2\pi^2{\boldsymbol\umag}\cdot{\bf M}{\boldsymbol\umag}},
\nonumber \\
\tilde J_{e_1} &=& -2\pi^2 R^2(u^2-v^2)F{\rm e}^{-2\pi {\rm i}{\boldsymbol x}_{\rm c}\cdot{\boldsymbol\umag}}{\rm e}^{-2\pi^2{\boldsymbol\umag}\cdot{\bf M}{\boldsymbol\umag}}, {\rm~~and}
\nonumber \\
\tilde J_{e_2} &=& -4\pi^2 R^2uv F {\rm e}^{-2\pi {\rm i}{\boldsymbol x}_{\rm c}\cdot{\boldsymbol\umag}}{\rm e}^{-2\pi^2{\boldsymbol\umag}\cdot{\bf M}{\boldsymbol\umag}}.
\end{eqnarray}
These lead to the inverse matrix of inner products (where $e^2=e_1^2+e_2^2$ and $\xi=1-e^2$, and we have used many applications of Wick's theorem):
\begin{eqnarray}
{\mathbfss H} \!\!\!\!&=& \!\!\!\! 4\pi R^2\sqrt{1-e^2}
\nonumber \\ && \!\!\!\!\!\!\!\!\!\!\!\!\!\!\!\times
\left( \!\begin{array}{cccccc}
2 & 0 & 0 & \frac RF & 0 & 0 \\
0 & \!\!\frac{2R^2(1+e_1)}{F^2}\!\! & \frac{2R^2e_2}{F^2} & 0 & 0 & 0 \\
0 & \frac{2R^2e_2}{F^2} & \!\!\frac{2R^2(1-e_1)}{F^2}\!\! & 0 & 0 & 0 \\
\frac RF & 0 & 0 & \frac{R^2(1+e^2)}{F^2} & \frac{2R\xi e_1}{F^2} & \frac{2R\xi e_2}{F^2} \\
0 & 0 & 0 & \frac{2R\xi e_1}{F^2} & \!\!\frac{4\xi(1-e_1^2)}{F^2}\!\! & \frac{-4\xi e_1e_2}{F^2} \\
0 & 0 & 0 & \frac{2R\xi e_2}{F^2} & \frac{-4\xi e_1e_2}{F^2} & \!\!\frac{4\xi(1-e_2^2)}{F^2}\!\! \\
\end{array}\!\right).
\nonumber \\ &&
\end{eqnarray}
From Eq.~(\ref{eq:NoiseQ.2}), the noise correlations projected into the data space can be written as an integral over the noise power spectrum:
\begin{equation}
{\mathbfss Q} = \int {\boldsymbol\tau}{\boldsymbol\tau}^\dagger e^{-4\pi^2{\boldsymbol\umag}\cdot{\mathbfss M}{\boldsymbol\umag}} P_N(u,v)\,{\rm d}u\,{\rm d}v,
\end{equation}
where the column vector ${\boldsymbol\tau}$ is a function of ${{\boldsymbol\umag}}$:
\begin{equation}
{\boldsymbol\tau} = \left( \begin{array}c
1 \\
-2\pi {\rm i}Fu \\
-2\pi {\rm i}Fv \\
-4\pi^2 RF[(1+e_1)u^2+2e_2uv+(1-e_1)v^2] \\
-2\pi^2 R^2F(u^2-v^2) \\
-4\pi^2 R^2Fuv \end{array} \right).
\end{equation}
We also need to use Eq.~(\ref{eq:JJ}) to evaluate $\langle J_{\gamma\delta}, J_{e_1} \rangle$ at zero ellipticity:
\begin{equation}
\langle J_{\gamma\delta}, J_{e_1} \rangle|_{e=0} = \frac1{8\pi R^2} \left( \begin{array}{cccccc}
0 & 0 & 0 & 0 & \frac F2 & 0 \\
0 & \frac{F^2}{2R^2} & 0 & 0 & 0 & 0 \\
0 & 0 & -\frac{F^2}{2R^2} & 0 & 0 & 0 \\
0 & 0 & 0 & 0 & -\frac{F^2}{2R} & 0 \\
\frac F2 & 0 & 0 & -\frac{F^2}{2R} & 0 & 0 \\
0 & 0 & 0 & 0 & 0 & 0 \\
\end{array} \right).
\end{equation}
After a tedious calculation, we find that the total ellipticity bias is
\begin{eqnarray}
\Delta e_1 \!\! &=&  \!\! - \frac{256\pi^4R^6}{F^2} \int [4\pi^2(u^2+v^2)R^2-1]^2 (u^2-v^2)
\nonumber \\ && ~\times {\rm e}
^{-4\pi^2R^2(u^2+v^2)}\,P_{\rm N}(u,v)\,{\rm d}u\,{\rm d}v.
\label{eq:DE1.ps}
\end{eqnarray}
An analogous equation holds for $\Delta e_2$:
\begin{eqnarray}
\Delta e_2 \!\! &=& \!\! - \frac{256\pi^4R^6}{F^2} \int [4\pi^2(u^2+v^2)R^2-1]^2 (2uv)
\nonumber \\ && ~\times
{\rm e}^{-4\pi^2R^2(u^2+v^2)}\,P_{\rm N}(u,v)\,{\rm d}u\,{\rm d}v.
\label{eq:DE2.ps}
\end{eqnarray}
The combined equation can be written in the form:
\begin{equation}
(\Delta e_1, \Delta e_2)
= -\frac{R^4}{F^2}\int W(\umag R) (\cos 2\varphi,\sin 2\varphi)
\,P_{\rm N}(u,v)\,{\rm d}u\,{\rm d}v,
\label{eq:DEcombo.ps}
\end{equation}
where $\varphi = \arctan (v/u)$ is the position angle of mode $(u,v)$, $P_{\rm N}(u,v)\,{\rm d}u\,{\rm d}v$ is the differential contribution to the noise intensity variance $\sigma_I^2$, and the function
\begin{equation}
W(z) = 256\pi^4 z^2(4\pi^2 z^2-1)^2{\rm e}^{-4\pi^2z^2}
\label{eq:Omega-u}
\end{equation}
is a dimensionless weight. The weight function goes to zero both at $z\rightarrow 0$ and $z\rightarrow\infty$: this is because correlated noise at zero spatial frequency is equivalent to an overall ``pedestal'' in the background at each galaxy, which may affect the flux measurement but does not bias the ellipticity. Noise at high spatial frequency $\umag\gtrsim 0.6/R$ is outside the band limit of the shape measurement, and hence has no effect.

\begin{figure}
\includegraphics[width=3.25in]{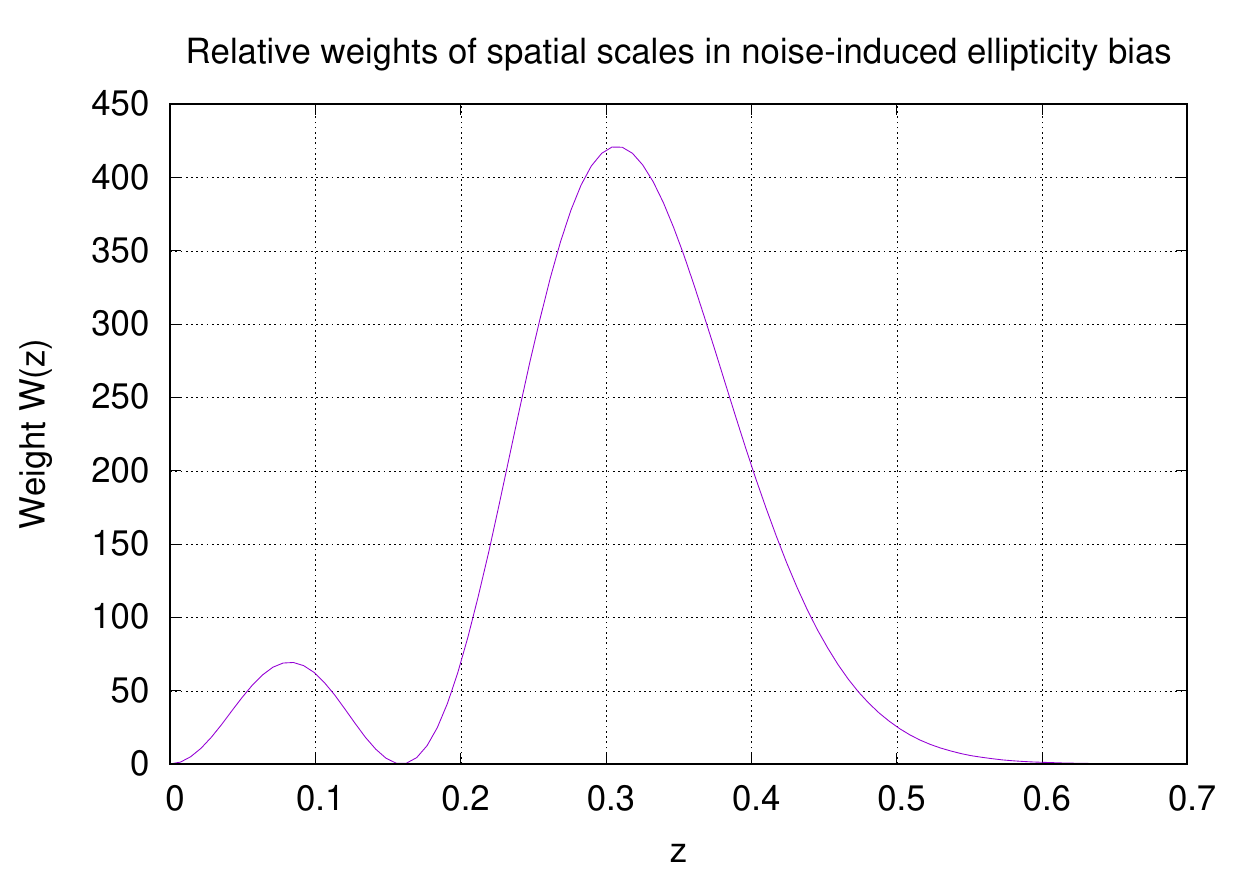}
\caption{\label{fig:weight}The weight function $W(z)$ in Eq.~(\ref{eq:DEcombo.ps}) as a function of wave number. Note that an anisotropic contribution to the noise is most damaging at a scale of $\umag=\sqrt{u^2+v^2} \approx 0.3/R$ (or a wavelength of $\umag^{-1}\approx 3 R$).}
\end{figure}

Note that the above considerations apply to the ``as-observed'' galaxy, e.g. ``$R$'' is the scale radius after convolution with the PSF, and $(\Delta e_1, \Delta e_2)$ is the noise-induced ellipticity bias before PSF correction. In e.g. a shape measurement method where the observed ellipticity is ``corrected'' by dividing by a responsivity factor ${\cal R}_2$ (usually between $\sim 0.3$ and 1), the bias on the reported ellipticity would also be divided by this factor. Also note that one must divide by 2 to get an ellipticity in the $g = (a-b)/(a+b)$ convention (where $a$ and $b$ are major and minor axes) instead of the $e=(a^2-b^2)/(a^2+b^2)$ convention:
\begin{equation}
(\Delta g_1, \Delta g_2)
= -\frac{R^4}{2F^2}\int W(\umag R) (\cos 2\varphi,\sin 2\varphi)
\,P_{\rm N}(u,v)\,{\rm d}u\,{\rm d}v.
\label{eq:noisebias-g}
\end{equation}

The analytic formula is useful since it is applicable to any shape of the noise power spectrum: weakly anisotropic noise, ``striping'' noise (correlated in the row direction), and at any spatial scale. This allows us to rapidly estimate the order of magnitude of the bias introduced by a noise source, and understand its scaling behavior. However it is also critical to compare the analytic model to simulations.

\subsection{Comparison to toy simulations}

\begin{figure}
\includegraphics[width=3in]{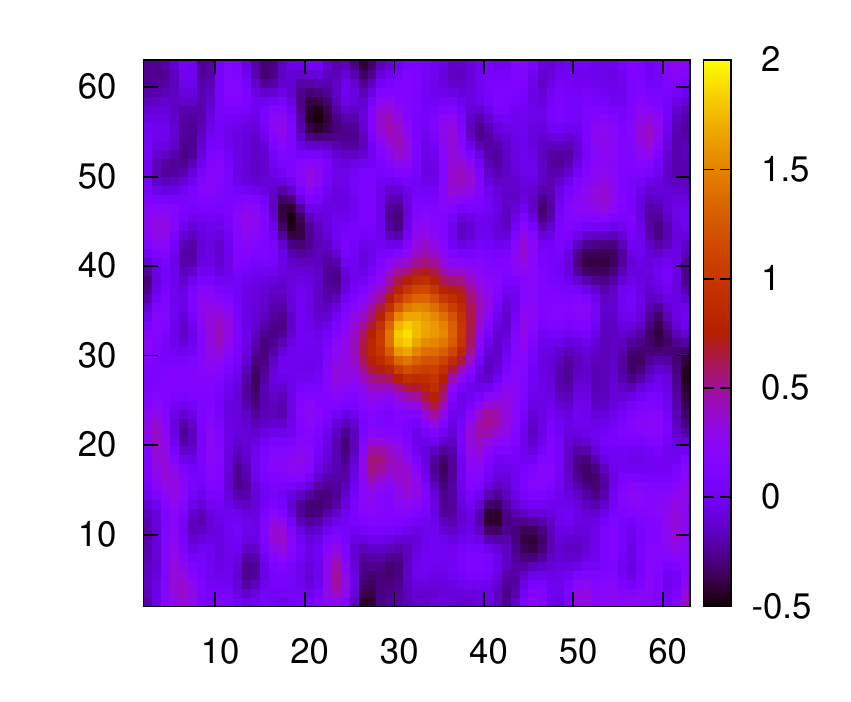}
\caption{\label{fig:map1_ex}An example circular galaxy with flux $F=200$ counts and scale radius $R=4$ pixels, superposed on the noise field of Eq.~(\ref{eq:NoiseCase1}) with $A=1$, $\alpha=4/\pi$, and $\beta=16/\pi$. That is, the correlation length of the noise is twice as large in the $y$-direction as the $x$-direction. Noise of this form leads to a bias of $e_1$ toward negative values, i.e., toward inferring a vertical orientation for the galaxy.}
\end{figure}

It is useful to compare the analytical bias estimate to a ``toy'' simulation designed to mimic the assumptions in the analytic model. This is a check of whether the analytical calculation is performed correctly, and also allows simple tests of its range of validity.

\begin{figure*}
\includegraphics[width=6.5in]{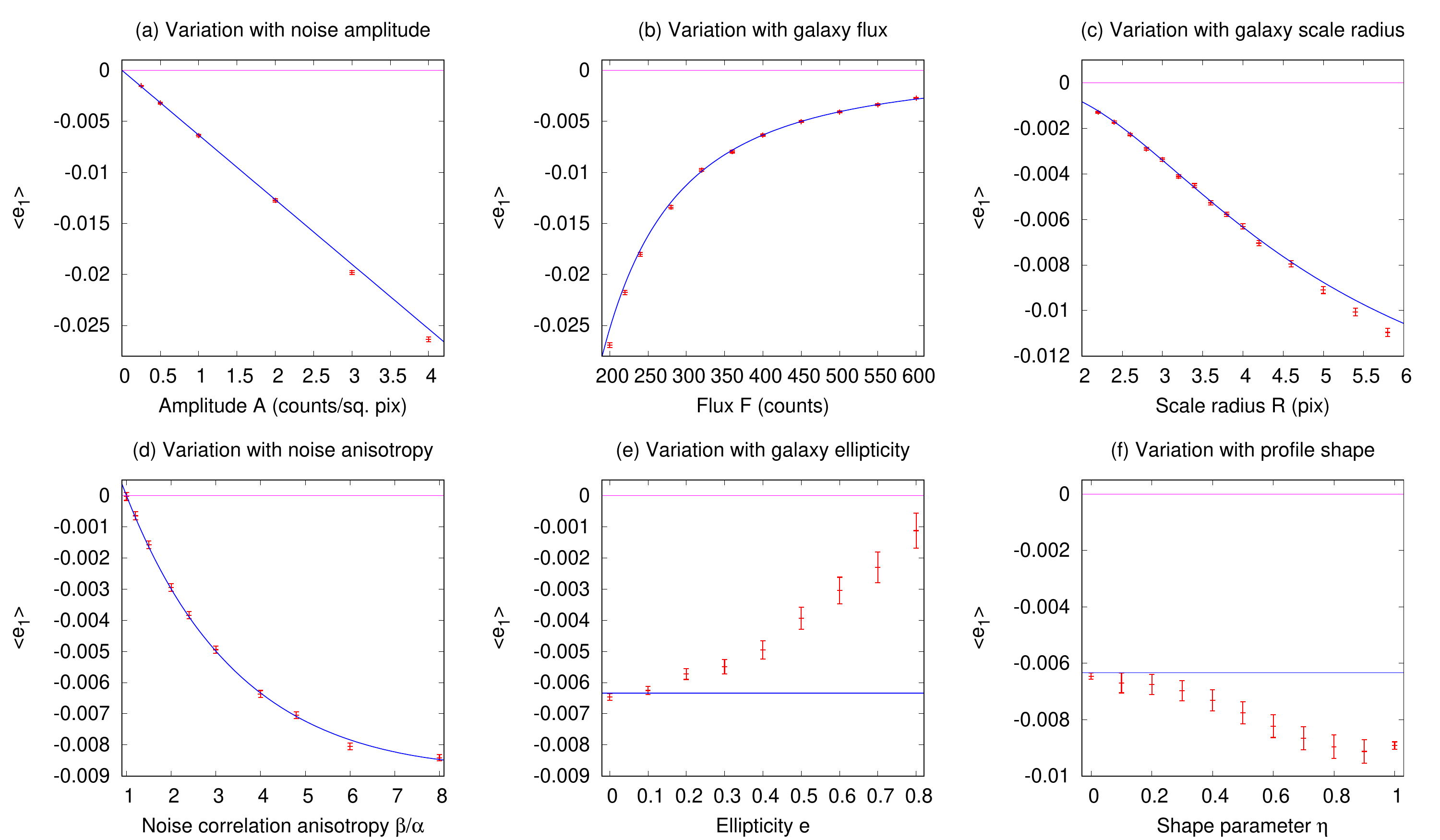}
\caption{\label{fig:biases}The shape measurement biases measured from simulations (red points with $1\sigma$ Monte Carlo errors) and calculated from the analytic approximation, Eq.~(\ref{eq:DE1.ps}) (blue lines). Each panel is based on the reference case (circular galaxy, $F=200$ counts, $R=4$ pixels, noise at $A=1$ with the noise spectrum of Eq.~(\ref{eq:NoiseCase1}) except for the variations indicated. In panels (a--d), the agreement is in general excellent, but for small signal-to-noise ratio, statistically significant deviations can be seen. Panels (e) and (f) show cases that violate the assumptions of the analytic model, and hence show larger errors. The number of simulations per point is $10^6$, except for the $0<\eta<1$ points in panel (f), which are more computationally expensive; in these cases $10^5$ simulations were run.}
\end{figure*}

The first set of such simulations is performed on a circular galaxy with flux $F=200$ counts, scale radius $R=4$ pixels, and centered at pixel $(32,32)$ of a $64\times 64$ simulation. A Gaussian random field was generated for the noise, with power spectrum:
\begin{equation}
P_{\rm N}(u,v) = A^2e^{-4\pi^2(\alpha u^2 - \beta v^2)},
\label{eq:NoiseCase1}
\end{equation}
where $A$ is an amplitude parameter (with units of counts per square pixel), and the default values of the constants are $\alpha = 4/\pi$ and $\beta=16/\pi$. This results in a noise field correlated in the $y$ direction; see Figure~\ref{fig:map1_ex}.

We may now investigate the biases in ellipticity resulting from this noise field. We generated $10^6$ random realizations of the noise, and measured the galaxy shape each time using the least-squares elliptical Gaussian method. We compare the ellipticity bias with the analytical prediction of Eq.~(\ref{eq:DE1.ps}) for a range of galaxy parameters. For the noise spectrum of Eq.~(\ref{eq:NoiseCase1}), the predicted bias can be obtained via integration over the Gaussian:
\begin{eqnarray}
\Delta e_1 \!\! &=& \!\! \frac{2\pi R^6 A^2(\alpha-\beta)}{F^2(\alpha+R^2)^{7/2}(\beta+R^2)^{7/2}}
\Bigl[
28R^8 + 20 (\alpha+\beta)R^6
\nonumber \\ && 
+ 7(\alpha-\beta)^2R^4 -4\alpha\beta(\alpha+\beta)R^2 + 4\alpha^2\beta^2
\Bigr],
\label{eq:degauss}
\end{eqnarray}
and this can be compared to the numerical results.

The outcome of these tests is shown in Figure~\ref{fig:biases}. We see that, as expected, the noise bias scales as the noise variance $\propto A^2$ (panel a) and as the inverse square of the galaxy flux, $\propto 1/F^2$ (panel b). The radius dependence is more complex (panel c) --- indeed, from Eq.~(\ref{eq:degauss}), one can see that the bias approaches a constant at large $R$, i.e. when the size of the galaxy is large compared to the correlation length of the noise. This is because for fixed flux, the reduction in signal-to-noise as the light is spread over more pixels cancels against the decreasing importance of the noise correlations (since they are over a small fraction of a galaxy scale length). The noise anisotropy ($\beta/\alpha$, or aspect ratio squared) is varied in panel (d), with fixed $\alpha$. Good agreement with the analytical model is seen in all of these cases.

Panels (e) and (f) of Fig.~\ref{fig:biases} show effects not treated in the analytical model. In panel (e), we have varied the ellipticity of the galaxies --- that is, instead of inserting a circular Gaussian galaxy, we have given it the indicated ellipticity $e=\sqrt{e_1^2+e_2^2}$ and assigned a random position angle $\chi = \frac12\arctan (e_2/e_1)$, and kept the same flux and scale radius $R$. It can be seen that for more elliptical galaxies, the bias becomes ``smaller'' (in absolute value, i.e. closer to zero).

Panel (f) shows the variation with the galaxy profile. For each value of $\eta$, a galaxy was generated via the convolution of an exponential profile and a Gaussian profile:
\begin{equation}
I = I_{\rm exp} \ast I_{\rm Gauss}, ~~~~I_{\rm exp}(r) \propto 2^{-r/r_{\rm e}},~~~~I_{\rm G}(r) \propto e^{-r^2/2r_{\rm G}^2},
\end{equation}
and with a ratio of radii $r_{\rm e}:r_{\rm G} = \eta:1-\eta$. This interpolates between the pure Gaussian limit ($\eta=0$) and pure exponential ($\eta=1$). This is intended to be a crude representation of a centrally peaked galaxy with some smearing. The normalization (amplitude) and total radius are set so that the least-squares Gaussian fit has the default values of $F=200$ and $R=4$.

\bsp	
\label{lastpage}
\end{document}